\pdfoutput=1
\documentclass[11pt,twoside,a4paper,cmspaper,final,collab]{cms-tdr}
\def\svnVersion{17319a6}\def\svnDate{2026/05/25}\def\cmsCernNoTag{CERN-EP-2025-116}\def\cmsCernDate{\today}\def\cmsMessage{Published in the Journal of High Energy Physics as \href{http://dx.doi.org/10.1007/JHEP05(2026)186}{\doi{10.1007/JHEP05(2026)186}.}}
\begin{document}\cmsNoteHeader{HIG-22-012}

\newcommand{\XZero}{\HepParticle{X}{}{(0)}\Xspace}
\newcommand{\XTwo}{\HepParticle{X}{}{(2)}\Xspace}
\newcommand{\PY}{\HepParticle{Y}{}{}\Xspace}

\newcommand{\mx}{\ensuremath{m_{\PX}}\xspace}
\newcommand{\my}{\ensuremath{m_{\PY}}\xspace}
\newcommand{\mgg}{\ensuremath{m_{\PGg\PGg}}\xspace}
\newcommand{\mtt}{\ensuremath{m_{\PGt\PGt}}\xspace}
\newcommand{\mH}{\ensuremath{m_{\PH}}\xspace}

\newcommand{\ttgammagamma}{\ensuremath{\ttbar + \PGg\PGg}\xspace}
\newcommand{\ttgamma}{\ensuremath{\ttbar + \PGg}\xspace}
\newcommand{\gjet}{\ensuremath{\PGg + \text{jets}}\xspace}
\newcommand{\dipho}{\ensuremath{\PGg\PGg + \text{jets}}\xspace}
\newcommand{\vgamma}{\ensuremath{\PV + \PGg}\xspace}

\newcommand{\DIPHO}{\ensuremath{\PGg\PGg+\text{jets}}\xspace}
\newcommand{\GAMJETS}{\ensuremath{\PGg+\text{jets}}\xspace}
\newcommand{\TTJETS}{\ensuremath{\ttbar + \text{jets}}\xspace}

\newcommand{\ttH}{\ensuremath{\ttbar\PH}\xspace}
\newcommand{\VH}{\ensuremath{\PV\PH}\xspace}
\newcommand{\VBF}{\ensuremath{\mathrm{VBF}}\xspace}
\newcommand{\ggH}{\ensuremath{\Pg\Pg\PH}\xspace}
\newcommand{\ggF}{\ensuremath{\Pg\Pg\mathrm{F}}\xspace}

\newcommand{\HH}{\ensuremath{\PH\PH}\xspace}
\newcommand{\YH}{\ensuremath{\PY\PH}\xspace}

\newcommand{\kl}{\ensuremath{\kappa_\lambda}\xspace}
\newcommand{\kapt}{\ensuremath{\kappa_\PQt}\xspace}
\newcommand{\lHHH}{\ensuremath{\lambda_{\PH\PH\PH}}\xspace}
\newcommand{\ctwo}{\ensuremath{c_2}\xspace}
\newcommand{\cg}{\ensuremath{c_{\Pg}}\xspace}
\newcommand{\cgg}{\ensuremath{c_{2\Pg}}\xspace}

\newcommand{\Htt}{\ensuremath{\PH \to \PGt\PGt}\xspace}
\newcommand{\XHH}{\ensuremath{\PX \to \HH}\xspace}

\newcommand{\XZeroHH}{\ensuremath{\XZero \to \HH}\xspace}
\newcommand{\XTwoHH}{\ensuremath{\XTwo \to \HH}\xspace}
\newcommand{\XYH}{\ensuremath{\PX \to \PY\PH}\xspace}
\newcommand{\XYggHtt}{\ensuremath{\PX \to \PY(\PGg\PGg)\PH(\PGt\PGt)}\xspace}
\newcommand{\XYttHgg}{\ensuremath{\PX \to \PY(\PGt\PGt)\PH(\PGg\PGg)}\xspace}
\newcommand{\XYbbHgg}{\ensuremath{\PX \to \PY(\PQb\PQb)\PH(\PGg\PGg)}\xspace}
\newcommand{\XYHHggtt}{\ensuremath{\PX \to (\PY/\PH)\PH \to \ggtt}\xspace}
\newcommand{\Ygg}{\ensuremath{\PY \to \PGg\PGg}\xspace}
\newcommand{\Ytt}{\ensuremath{\PY \to \PGt\PGt}\xspace}
\newcommand{\ggtt}{\ensuremath{\PGg\PGg\PGt\PGt}\xspace}
\newcommand{\HHggtt}{\ensuremath{\PH\PH \to \ggtt}\xspace}

\newcommand{\TwoNLL}{\ensuremath{-2\Delta\ln{L}}\xspace}

\cmsNoteHeader{HIG-22-012}
\title{Search for the nonresonant and resonant production of a Higgs boson in association with an additional scalar boson in the \texorpdfstring{\ggtt}{gamma-gamma-tau-tau} final state in proton-proton collisions at \texorpdfstring{$\sqrt{s} = 13\TeV$}{sqrt(s) = 13 TeV}}

\date{\today}

\abstract{
The results of a search for the production of two scalar bosons in final states with two photons and two tau leptons are presented. The search considers both nonresonant production of a Higgs boson pair, \HH, and resonant production via a new boson \PX which decays either to \HH or to \PH and a new scalar \PY. The analysis uses up to 138\fbinv of proton-proton collision data, recorded between 2016 and 2018 by the CMS experiment at the LHC at a center-of-mass energy of 13\TeV. No evidence for signal is found in the data. For the nonresonant production, the observed (expected) upper limit at 95\% confidence level (\CL) on the \HH production cross section is set at 930 (740)\unit{fb}, corresponding to 33 (26) times the standard model prediction. At 95\% \CL, \HH production is observed (expected) to be excluded for values of \kl outside the range between $-12$ ($-9.4$) and 17 (15). Observed (expected) upper limits at 95\% \CL for the \XHH cross section are found to be within 160 to 2200 (200 to 1800)\unit{fb}, depending on the mass of \PX. In the \XYttHgg search, the observed (expected) upper limits on the product of the production cross section and decay branching fractions vary between 0.059--1.2\unit{fb} (0.087--0.68\unit{fb}). For the \XYggHtt search the observed (expected) upper limits on the product of the production cross section and \Ygg branching fraction vary between 0.69--15\unit{fb} (0.73--8.3\unit{fb}) in the low \PY mass search, tightening constraints on the next-to-minimal supersymmetric standard model, and between 0.64--10\unit{fb} (0.70--7.6\unit{fb}) in the high \PY mass search.
}

\hypersetup{%
pdfauthor={CMS Collaboration},%
pdftitle={Search for the nonresonant and resonant production of a Higgs boson in association with an additional scalar boson in the ggtt final state in proton-proton collisions at sqrt(s) = 13 TeV},%
pdfsubject={CMS},%
pdfkeywords={CMS, dihiggs, nonresonant, resonant, photons, taus}}

\maketitle

\section{Introduction\label{sec:intro}}
The Higgs boson (\PH) holds a unique position in the standard model (SM) of particle physics as the only fundamental scalar boson. Since its discovery in 2012 at the CERN LHC by the ATLAS and CMS Collaborations~\cite{ATLAS:2012yve, CMS:2012qbp, CMS:2013btf}, there has been an extensive research program to characterize its properties. So far, the properties have been measured to be consistent with SM predictions~\cite{CMS:2022dwd, NatureHiggsATLAS2022}. Nevertheless, the measurements of several Higgs boson properties are extremely challenging at the LHC, leading to relatively weak constraints. One of the least constrained properties is the strength of the Higgs trilinear self-interaction, \lHHH. Within the SM, the value of \lHHH and subsequently the structure of the Higgs field potential are completely determined, given the Higgs boson mass, \mH, and the vacuum expectation value of the Higgs field. Measurements of \lHHH thus provide a unique consistency test of the SM and impose constraints on the shape of the scalar potential.

\begin{figure}
    \centering
    \includegraphics[width=.4\textwidth]{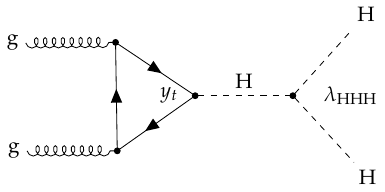}\hspace{0.1\textwidth}
    \includegraphics[width=.4\textwidth]{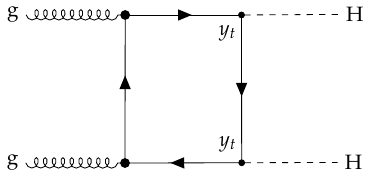}\\
    \includegraphics[width=.3\textwidth]{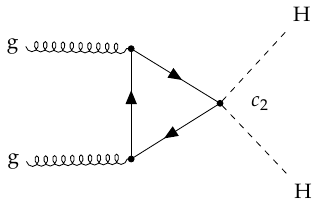}\hspace{0.03\textwidth}
    \includegraphics[width=.3\textwidth]{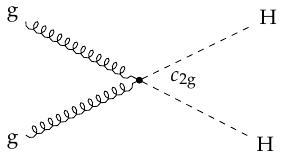}\hspace{0.03\textwidth}
    \includegraphics[width=.3\textwidth]{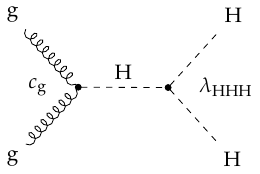}
    \caption{Leading order Feynman diagrams of the nonresonant \HH production via \ggF. The two diagrams in the upper row correspond to SM processes, involving the top Yukawa coupling $y_\cPqt$, and the trilinear Higgs boson self-coupling \lHHH. The diagrams in the lower row correspond to BSM processes involving contact interactions introduced in the effective field theory, namely $\ctwo$, $\cgg$ and $\cg$. \label{fig:feynman_diagrams_nonres}}
\end{figure}

At the LHC, \lHHH is directly constrained through measurements of Higgs boson pair (\HH) production. The primary production mechanism is via gluon-gluon fusion (\ggF), with representative leading order (LO) diagrams shown in Fig.~\ref{fig:feynman_diagrams_nonres}~(upper row). The destructive interference between diagrams leads to a small SM cross section at $\sqrt{s}=13$\TeV of $31.1^{+2.1}_{-7.2}$\unit{fb}, calculated for $\mH = 125\GeV$ at next-to-next-to-leading order (NNLO), with resummation at next-to-next-to-leading-logarithm accuracy, and scale and top quark mass effects at next-to-leading order (NLO)~\cite{Grazzini:2018bsd,Dawson:1998py,Borowka:2016ehy,Baglio:2018lrj,deFlorian:2013jea,Shao:2013bz,deFlorian:2015moa,Baglio:2020wgt}. This value is three orders of magnitude smaller than the single \PH production cross section, explaining why \HH production is yet to be observed experimentally. The ultimate sensitivity is achieved via the combined analysis of final states, where CMS and ATLAS have recently reported upper limits on the \HH cross section of 3.4 and 2.4 times the SM prediction at 95\% confidence level (\CL), respectively~\cite{CMS:2022dwd,ATLAS:2022jtk}.

The cross section and kinematical distributions of HH production can be significantly affected in many theories of physics beyond-the-SM (BSM). If the new particles introduced in these theories have sufficiently high masses, they will not be produced at current LHC energies. This motivates the use of an effective field theory (EFT) framework to parameterize the contributions of new physics in terms of higher-dimensional operators, suppressed by a large mass scale $\Lambda$.
In this analysis, BSM contributions to nonresonant \ggF \HH production are parameterized through an effective Lagrangian that extends that of the SM with dimension-6 operators~\cite{LHCHiggsCrossSectionWorkingGroup:2016ypw,Goertz:2014qta}. This effective Lagrangian parameterizes BSM effects in terms of five couplings involving the Higgs boson. Two are present in the SM and are expressed as ratios to their SM values: the trilinear self-coupling ($\kl = \lHHH/\lHHH^{\text{SM}}$) and the top quark Yukawa coupling ($\kapt = y_\PQt/y_\PQt^{\text{SM}}$), while three are contact interactions not present in the SM: between two Higgs bosons and two top quarks (\ctwo), between two Higgs bosons and two gluons (\cgg), and between one Higgs boson and two gluons (\cg). Diagrams involving each of these five couplings are shown in Fig.~\ref{fig:feynman_diagrams_nonres} (lower row).

Moreover, many BSM theories predict the existence of additional scalar particles whose production can lead to the same final states studied in \HH production. For example, the Randall--Sundrum (RS) bulk model~\cite{Randall:1999ee,Goldberger:1999uk,Davoudiasl:1999jd}, which predicts the existence of a spin-0 resonance referred to as the radion, and a spin-2 first Kaluza--Klein excitation of the graviton. Both resonances can decay to \HH with a sizeable branching fraction, thus motivating the search for the resonant production of a Higgs boson pair.

\begin{figure}
    \centering
    \includegraphics[width=0.3\textwidth]{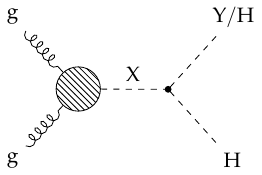}
    \caption{Feynman diagram of the resonant production of a pair of SM Higgs bosons (\XHH), or of a SM Higgs boson and a new scalar particle (\XYH).\label{fig:feynman_diagrams_res}}
\end{figure}

Another example of a BSM theory is supersymmetry (SUSY), which extends upon the SM by introducing a fermionic (bosonic) partner for each boson (fermion) in the SM. One such theory, known as the next-to-minimal supersymmetric SM (NMSSM)~\cite{Djouadi:2005gj}, introduces an additional complex Higgs doublet and a complex singlet field. This scenario amounts to five neutral Higgs bosons, one of which being associated with the 125\GeV Higgs boson, \PH. Within the NMSSM, the heavier Higgs boson \PX may decay to lighter Higgs bosons (\PH or \PY), inspiring the search for the \XYH process at the LHC. A Feynman diagram for the resonant production of two scalar bosons is shown in Fig.~\ref{fig:feynman_diagrams_res}.

Recent results from the CMS experiment using data collected between 2016 and 2018 have yielded excesses in specific final states. An excess with a local (global) significance of 3.8 (2.8) standard deviations was observed in the \XYbbHgg search at $\mx=650\GeV$ and $\my=90\GeV$~\cite{CMS:2023boe}. The search for a scalar particle decaying to a tau lepton pair yielded excesses at scalar masses of around 0.1 and 1.2\TeV with local (global) significances of 3.1 (2.7) and 2.8 (2.2) standard deviations respectively~\cite{CMS:2022goy}. In addition, using data collected between 2016 and 2018, a search for a light Higgs boson decaying to a photon pair observed an excess with a local (global) significance of 2.9 (1.3) standard deviations at 95.4\GeV~\cite{CMS:2024yhz}. The compatibility in mass across these channels motivates searching for new scalar particles in different final states, complementing previous searches involving photons and tau leptons.

This paper presents a number of searches for the production of two scalar bosons in the \ggtt final state using proton-proton (pp) collision data collected by the CMS experiment between 2016 and 2018 at $\sqrt{s}=13$\TeV, corresponding to an integrated luminosity of 138 or 132\fbinv, depending on the trigger path. Despite the small SM \HHggtt branching fraction, the diphoton pair offers a clean experimental signature to trigger on with a good mass resolution, whilst the additional tau leptons in the event help further isolate signal from background. In total, there are five searches:

\begin{enumerate}
    \item Nonresonant \HH production via \ggF.
    \item Resonant \HH production via a spin-0 particle, \XZeroHH.
    \item Resonant \HH production via a spin-2 particle, \XTwoHH.
    \item Resonant \XYH production, in the \XYttHgg decay channel.
    \item Resonant \XYH production, in the \XYggHtt decay channel. 
\end{enumerate}

In addition to a search for \HH production, the nonresonant search is used to place direct constraints on \kl, under the assumption that all other couplings of the Higgs boson have values as predicted by the SM. This search is also used to constrain EFT parameters by placing limits on the cross section of \ggF \HH production for thirteen non-SM benchmark scenarios involving the five couplings: $\kl$, $\kapt$, $\cgg$, $\cg$, and $\ctwo$. The benchmark scenarios are those proposed in Refs.~\cite{Carvalho:2015ttv, Buchalla:2018yce}.

In the resonant searches, no assumptions are made regarding the specific BSM theory to which new particles \PX and \PY belong other than that they have a narrow width, are spin-0 except for \XTwo which is spin-2, and that \PY can decay to two photons or two tau leptons. The \XZeroHH and \XTwoHH searches consider the mass of \PX to be in the range from 260 to 1000\GeV. The \XYttHgg search considers a range in \mx from 300 to 1000\GeV and a range in \my from 50 to 800\GeV, whilst the \XYggHtt search considers the same \mx range, with a slightly reduced range in \my, from 70 to 800\GeV, which is imposed by a triggering requirement of $\mgg>65\GeV$. Since \PX is the mother particle and \PY is a decay product, \mx is required to be greater than $\my+\mH$.

In this analysis, results are extracted by performing fits to the diphoton invariant mass distribution, \mgg, in event categories designed to be enriched in signal events from one of the five searches. The first four searches amount to fitting a peak at \mH~since the diphoton system is produced via the decay of a SM Higgs boson, \HGG. In the \XYggHtt search, the strategy is different as the peak formed by \Ygg events must be fit over the wide range of considered \my values. To account for variations in the trigger efficiencies and event kinematics for different mass regimes, the \XYggHtt search is effectively split into two searches.

\begin{enumerate}
    \item Low-mass \XYggHtt search considers \my in the range from 70 to 125\GeV.
    \item High-mass \XYggHtt search considers \my in the range from 125 to 800\GeV.
\end{enumerate}

This paper is organized as follows: Section~\ref{sec:detector} gives a description of the CMS detector and of the event reconstruction. The data and simulations used are detailed in Section~\ref{sec:samples}. An overview of the analysis strategy is provided in Section~\ref{sec:strategy}, then the event selection is described in Section~\ref{sec:event_reco}, followed by Sections~\ref{sec:nonres_event_cat} and~\ref{sec:res_event_cat}, which explain the event categorization procedure for the nonresonant and resonant search, respectively. The signal and background modeling is described in Section~\ref{sec:sig_bkg_modeling}, and the treatment of systematic uncertainties is detailed in Section~\ref{sec:systematics}. The results for each search are presented in Section~\ref{sec:results}, and a summary of the paper is provided in Section~\ref{sec:summary}. Tabulated results are provided in the HEPData record for this analysis~\cite{hepdata}.

\section{The CMS detector and event reconstruction\label{sec:detector}}
The central feature of the CMS apparatus is a superconducting solenoid of 6\unit{m} internal diameter, providing a magnetic field of 3.8\unit{T}. Within the solenoid volume are a silicon pixel and strip tracker, a lead tungstate crystal electromagnetic calorimeter (ECAL), and a brass and scintillator hadron calorimeter (HCAL), each composed of a barrel and two endcap sections. Forward calorimeters extend the pseudorapidity ($\eta$) coverage provided by the barrel and endcap detectors. Muons are measured in gas-ionization detectors embedded in the steel flux-return yoke outside the solenoid. A more detailed description of the CMS detector, together with a definition of the coordinate system used and the relevant kinematical variables, can be found in Ref.~\cite{CMS:2008xjf}.

Events of interest are selected using a two-tiered trigger system. The first level (L1), composed of custom hardware processors, uses information from the calorimeters and muon detectors to select events at a rate of around 100\unit{kHz} within a fixed latency of 4\mus~\cite{CMS:2020cmk}. The second level, known as the high-level trigger (HLT), consists of a farm of processors running a version of the full event reconstruction software optimized for fast processing, and reduces the event rate to around 1\unit{kHz} before data storage~\cite{CMS:2016ngn,CMS:2024aqx}.

A particle-flow (PF) algorithm~\cite{CMS:2017yfk} aims to reconstruct and identify each individual particle in an event, with an optimized combination of information from the various elements of the CMS detector. Photons and electrons are reconstructed, at least in part, from energy deposits in the ECAL. Reconstruction begins with a clustering algorithm that groups ECAL crystals with energies exceeding a given threshold. Then, clusters within a certain geometric area are merged into ``superclusters'' to include photon conversions and bremsstrahlung losses. Photon candidates are identified by clusters in the ECAL not linked to any charged-particle trajectories seeded in the pixel detector. The energy of the photon is then estimated by summing the energy of each crystal in the supercluster, calibrated and corrected for response variations in time. The direction of the photon is determined by the positions of the primary vertex (PV) and the supercluster. The PV is taken to be the vertex corresponding to the hardest scattering in the event, evaluated using tracking information alone, as described in Section 9.4.1 of Ref.~\cite{CMS-TDR-15-02}. An electron is identified by an ECAL cluster that is linked to a charged-particle trajectory, and its energy is estimated from a combination of the energy of the ECAL supercluster, and the momentum of the electron at the PV as determined by the tracker. Further detail about these reconstruction algorithms can be found in Ref.~\cite{CMS:2020uim}.

The energy scale and resolution of photons and electrons are corrected in two stages. First, a multivariate regression technique based on simulation alone is employed to account for energy not being entirely deposited inside operational ECAL crystals. Then, a further set of corrections is derived using $\PZ \to \Pe\Pe$ events, such that simulation matches data. Both of these procedures are described in Ref.~\cite{CMS:2020uim}.

The energy of muons is obtained from the curvature of the corresponding track. The energy of charged hadrons is determined from a combination of their momentum measured in the tracker and the matching ECAL and HCAL energy deposits, corrected for the response function of the calorimeters to hadronic showers. Finally, the energy of neutral hadrons is obtained from the corresponding corrected ECAL and HCAL energy deposits.

In the barrel section of the ECAL, an energy resolution of about 1\% is achieved for unconverted or late-converting photons in the tens of\GeV energy range. The energy resolution of the remaining barrel photons is about 1.3\% up to $\abs{\eta} = 1$, changing to about 2.5\% at $\abs{\eta} = 1.4$. In the endcaps, the energy resolution is about 2.5\% for unconverted or late-converting photons, and between 3 and 4\% for the other ones~\cite{CMS:2015myp}. The diphoton mass resolution, as measured in \HGG~decays, is typically in the 1--2\% range, depending on the measurement of the photon energies in the ECAL and the topology of the photons in the event~\cite{CMS:2020xrn}.

The electron momentum is estimated by combining the energy measurement in the ECAL with the momentum measurement in the tracker. The momentum resolution for electrons with transverse momentum ($\pt$) equal to about $45\GeV$ from $\PZ \to \Pe \Pe$ decays ranges from 1.6 to 5\%. It is generally better in the barrel region than in the endcaps, and also depends on the bremsstrahlung energy emitted by the electron as it traverses the material in front of the ECAL~\cite{CMS:2020uim,CMS-DP-2020-021}.

Muons are measured in the pseudorapidity range $\abs{\eta} < 2.4$, with detection planes made using three technologies: drift tubes, cathode strip chambers, and resistive plate chambers. The single muon trigger efficiency exceeds 90\% over the full $\eta$ range, and the efficiency to reconstruct and identify muons is greater than 96\%. Matching muons to tracks measured in the silicon tracker results in a relative transverse momentum resolution, for muons with \pt up to 100\GeV, of 1\% in the barrel and 3\% in the endcaps. The \pt resolution in the barrel is better than 7\% for muons with \pt up to 1\TeV~\cite{CMS:2018rym}.

For each event, hadronic jets are clustered from these reconstructed particles using the infrared- and collinear-safe anti-\kt algorithm~\cite{Cacciari:2008gp, Cacciari:2011ma} with a distance parameter of 0.4.
Jet momentum is determined as the vectorial sum of all particle momenta in the jet, and is found from simulation to be, on average, within 5 to 10\% of the true momentum over the whole \pt spectrum and detector acceptance. Additional pp interactions within the same or nearby bunch crossings (pileup) can contribute additional tracks and calorimetric energy depositions, increasing the apparent jet momentum. To mitigate this effect, tracks identified to be originating from pileup vertices are discarded and an offset correction is applied to correct for remaining contributions~\cite{CMS:2020ebo}. Jet energy corrections are derived from simulation studies so that the average measured energy of jets becomes identical to that of particle level jets. In situ measurements of the momentum balance in dijet, $\text{photon} + \text{jet}$, $\PZ + \text{jet}$, and multijet events are used to determine any residual differences between the jet energy scale in data and in simulation, and appropriate corrections are made~\cite{CMS:2016lmd}. Additional selection criteria are applied to each jet to remove jets potentially dominated by instrumental effects or reconstruction failures~\cite{CMS:2020ebo}. Jets originating from b quarks are identified using the DEEPJET algorithm~\cite{CMS:2017wtu,Bols:2020bkb,CMS-DP-2018-058} at the medium working point defined by a light-quark or gluon jet misidentification rate of 1\%, and a b jet identification efficiency of 75\%.

Hadronic \PGt lepton decays (\tauh) are reconstructed from jets, using the hadrons-plus-strips algorithm~\cite{CMS:2018jrd}, which combines 1 or 3 tracks with energy deposits in the calorimeters, to identify the tau decay modes. Neutral pions are reconstructed as strips with dynamic size in $\eta$-$\phi$ from reconstructed electrons and photons, where the strip size varies as a function of the $\pt$ of the electron or photon candidate. A correction to the \tauh energy scale in simulation is derived by comparing simulation and data in $\PZ/\PGg^*$ events~\cite{CMS:2018jrd}.

To distinguish genuine \tauh decays from jets originating from the hadronization of quarks or gluons, and from electrons, or muons, the \textsc{DeepTau} algorithm is used~\cite{CMS:2022prd}. Information from all individual reconstructed particles near the \tauh axis is combined with properties of the $\tauh$ candidate and the event. The algorithm provides three discriminants, $D_\Pe$, $D_\PGm$, and $D_\text{jet}$, which are designed to distinguish \tauh candidates from electrons, muons, and jets, respectively. 
The efficiency for identifying a genuine \tauh candidate using these discriminants is measured in data with $\Z \to \PGt\PGt \to \tauh \PGm$ events for taus with $\pt < 140\GeV$ and $\PW^* \to \PGt\PGnGt$ events for $\pt > 140\GeV$. Corrections are applied to simulated events, such that the efficiencies measured in simulation match those measured in data.

The rate of a jet to be misidentified as \tauh by the \textsc{DeepTau} algorithm depends on the \pt and quark flavor of the jet. In simulated events from \PW boson production in association with jets, it has been estimated to be 0.43\% for a genuine \tauh identification efficiency of 70\%. The misidentification rate for electrons (muons) is 2.60 (0.03)\% for a genuine \tauh identification efficiency of 80 ($>$99)\%.

The missing transverse momentum vector \ptvecmiss is computed as the negative vector sum of the transverse momenta of all the PF candidates in an event, and its magnitude is denoted as \ptmiss~\cite{CMS:2019ctu}. The \ptvecmiss is modified to account for corrections to the energy scale of the reconstructed jets in the event.

\section{Data and simulated samples\label{sec:samples}}
This analysis uses data collected by the CMS detector during LHC pp collisions corresponding to a total integrated luminosity of 138\fbinv, in which 36.3, 41.5, and 59.8\fbinv were collected during 2016, 2017, and 2018, respectively~\cite{CMS:2021xjt,CMS-PAS-LUM-17-004,CMS-PAS-LUM-18-002}. In 2018, the trigger used in the low-mass \XYggHtt search was not introduced until after the start of that year's data taking, resulting in a lower integrated luminosity of 54.4\fbinv.

The Monte Carlo (MC) simulation of the ggF \HH signal is generated with \POWHEG 2.0 using the full top quark mass dependence~\cite{Buchalla:2018yce} and with NLO accuracy in perturbative quantum chromodynamics (QCD)~\cite{Bagnaschi:2011tu,Heinrich:2017kxx,Heinrich:2019bkc,Jones:2017giv,Heinrich:2020ckp,Nason:2004rx,Frixione:2007vw,Alioli:2010xd}. Both SM ($\kl=1$) and BSM ($\kl=0, 2.45, 5$) scenarios are studied. The dependence of the \ggF \HH cross section on \kl and \kapt can be expressed, to any order in QCD, by three terms, corresponding to diagrams involving \kl, \kapt, and their interference~\cite{Heinrich:2019bkc}. Therefore, any linear superposition of three samples involving unique \kl values can be used to obtain the signal sample corresponding to an arbitrary point in the $(\kl, \kapt)$ parameter space.

Following the recommendations from Refs.~\cite{Carvalho:2015ttv,Buchalla:2018yce}, a reweighting of the NLO \kl samples is applied to produce 13 EFT benchmark hypotheses, each corresponding to a different point in the 5D parameter space defined by the couplings: \kl, \kapt, \ctwo, \cg, and \cgg. 
The list of benchmark hypotheses generated is provided in Table~\ref{tab:bench}. 
The reweighting procedure operates at the event level by using a parameterization of the differential cross section in the true invariant mass of the HH system and the angular distance between the two Higgs bosons in the azimuthal plane, which completely characterizes the hard scattering because it only has two degrees of freedom~\cite{deFlorian:2021azd,Carvalho:2017vnu}.

\begin{table}
  \setlength{\tabcolsep}{3pt}
  \centering
  \topcaption{Parameter values of nonresonant BSM benchmark hypotheses. The first column corresponds to the SM sample, while the next 13 correspond to the benchmark hypotheses identified using the method from Refs.~\cite{Carvalho:2015ttv,Buchalla:2018yce}. \label{tab:bench}}
  \newcolumntype{R}{>{$}{r}<{$}}
  \begin{tabular}{l  R R R   R R R   R R R  R R R  R R}
  &     \text{SM} & 1 &   2 &    3 &    4 &   5 &   6 &   7 &    8 & 8a &   9 &   10 &  11 &   12 \\
  \hline
  $\kl$  & 1.0  &  7.5 &  1.0 &  1.0 & -3.5 &  1.0 &  2.4 &  5.0 & 15.0 & 1.0 &  1.0 & 10.0 &  2.4 & 15.0 \\
  $\kapt$  & 1.0  &  1.0 &  1.0 &  1.0 &  1.5 &  1.0 &  1.0 &  1.0 & 1.0 & 1.0 &  1.0 &  1.5 &  1.0 &  1.0 \\
  $\ctwo$  & 0.0  & -1.0 &  0.5 & -1.5 & -3.0 &  0.0 &  0.0 &  0.0 & 0.0 & 0.5 &  1.0 & -1.0 &  0.0 &  1.0 \\
  $\cg$    & 0.0  &  0.0 & -0.8 &  0.0 &  0.0 &  0.8 &  0.2 &  0.2 & -1.0 & 4/15 & -0.6 &  0.0 &  1.0 &  0.0 \\
  $\cgg$   & 0.0  &  0.0 &  0.6 & -0.8 &  0.0 & -1.0 & -0.2 & -0.2 & 1.0 &  0.0 &  0.6 &  0.0 & -1.0 &  0.0 \\
  \end{tabular}
\end{table}

In the resonant searches, we consider the production of \PX via ggF. The signal simulation samples for these searches are generated using \MGvATNLO (version 2.6.5) at LO accuracy~\cite{MadGraph}. Within the search ranges of \mx and \my, 17, 17, 91 and 83 simulation samples are generated for the \XZeroHH, \XTwoHH, \XYttHgg and \XYggHtt processes respectively. 

The dominant irreducible background for this analysis arises from prompt diphoton production ($\DIPHO$), while a dominant reducible background comes from $\GAMJETS$ events, where jets can be misidentified as photons or tau leptons. The $\DIPHO$ process is modeled at LO with \SHERPA v.2.2.1~\cite{10.21468/SciPostPhys.7.3.034}, which includes tree-level processes with up to three additional jets, as well as box processes at LO accuracy. The $\GAMJETS$ process is instead modeled at LO with \MGvATNLO v2.6.5~\cite{MadGraph, MadGraphMLM}.

Other nonresonant backgrounds include \vgamma, \ttbar, \ttgamma, and \ttgammagamma where \ttbar is a top quark-antiquark pair and the vector bosons (either from direct production or decays of the top quark) can decay into \PGt leptons and associated jets can be misidentified as photons. These processes are simulated at NLO in QCD using \MGvATNLO v2.6.5~\cite{MadGraph, MadGraphMLM,MadGraphFXFX} for all except \ttbar which uses \MGvATNLO v2.6.1.

Single Higgs boson production in the SM, where the Higgs boson decays to a pair of photons, is considered a resonant background. We include the following production modes: ggF, vector-boson fusion (\VBF), vector boson associated production (\VH), and production associated with a \ttbar pair (\ttH). These processes are simulated using \MGvATNLO v2.4.2, with cross sections and decay branching ratios taken from Ref.~\cite{LHCHiggsCrossSectionWorkingGroup:2016ypw}. No other processes with a single Higgs boson are considered, as their contributions are negligible.

All simulated samples are interfaced using \PYTHIA 8.240 with the CP5 tune~\cite{CMS:2015wcf, CMS:2019csb} for parton showering, fragmentation with the standard \pt-ordered parton shower scheme and the underlying event description. Parton distribution functions (PDFs) are taken from the NNLO NNPDF 3.1 set~\cite{NNPDF31}. The detector response is modeled using the \GEANTfour package~\cite{Agostinelli:2002hh}. The simulation includes a description of pileup interactions and a reweighting is applied to the simulated events such that the distribution of the number of interaction vertices matches that seen in data.

\section{Analysis strategy\label{sec:strategy}}
Proton-proton collision events deemed compatible with the \ggtt final state are selected starting from a diphoton high-level trigger with asymmetric photon \pt thresholds of 30 and 18\GeV. For all except the low-mass \XYggHtt search, the subleading photon threshold is increased to 22\GeV for 2017 and 2018. The triggers also apply loose isolation requirements. After the trigger selection, the photon candidates are required to pass a loose selection on kinematical variables, with different thresholds for the low-mass \XYggHtt, and a photon identification criterion, as discussed in Section~\ref{sec:event_reco}. The invariant mass of the diphoton candidate, \mgg, is required to be greater than 65\GeV for the low-mass \XYggHtt search, and greater than 100\GeV for all other searches.
In addition, to enter the analysis events must contain a valid $\PGt\PGt$ candidate or have a single \tauh candidate, as detailed in Section~\ref{sec:event_reco}.

In each search, a separate machine-learning classifier is used to further isolate events with signal-like characteristics from background. The classifiers are trained independently in each search using MC simulation of signal and background events. For the nonresonant search, a boosted decision tree (BDT) discriminant is applied which uses input features related to the kinematical properties of the event constituents. Sequential boundaries are placed on the BDT output to define event categories of different signal purity, where the boundary positions are chosen to maximize signal sensitivity. Further details are provided in Section \ref{sec:nonres_event_cat}.

In the resonant searches, we look for signal events produced over a wide range of resonance mass hypotheses. As the kinematical properties of the signal events vary dramatically over the considered mass range, it would be suboptimal to train a single classification algorithm using all mass points to identify signal from background. However, it would not be feasible to train a separate classification algorithm at each mass point due to the large number of mass points considered. A parameterized neural network (pNN)~\cite{Baldi:2016fzo} is thus employed, which uses the nominal \mx (and \my for the \XYH searches) as additional training features.
The pNN is trained with signal simulation from multiple mass hypotheses simultaneously, allowing it to implicitly learn how the kinematical distributions evolve with \mx (and \my) and thereby change the way it discriminates between signal and background based upon the resonance mass. In this approach, only one pNN is needed for each of the resonant searches, leading to five pNNs in total. Furthermore, provided that enough mass hypotheses within a given range are used in training, the pNN can be applied to mass hypotheses inside that range, regardless of whether they were used in training or not.

A set of analysis categories is defined by placing boundaries on the pNN output, where the pNN is evaluated at the values of \mx (and \my) that correspond to the mass hypothesis considered. The boundary positions are chosen to provide good signal sensitivity whilst maintaining a sufficient background yield to be able to fit the background model successfully. A new set of categories with its own boundaries is defined using the pNN output for each mass hypothesis considered.

The granularity in \mx (and \my) chosen to evaluate the pNN and define the analysis categories is motivated by the experimental resolution in \mx (and \my). Initially, a granularity of mass hypotheses corresponding to the simulation samples generated is considered, and for each hypothesis, results are obtained using categories defined for that hypothesis and also for neighbouring hypotheses in \mx (and \my). If the results differ by more than 10\%, the event selection is considered to have changed significantly enough to justify a finer granularity. In this case, the ranges between the mass hypotheses are divided into equal-sized intervals to keep the variations of the results below 10\%, under the assumption that the results vary linearly. In total, there are 27, 27, 398, 1765 and 2547 mass hypotheses considered in the \XZeroHH, \XTwoHH, \XYttHgg, low-mass \XYggHtt and high-mass \XYggHtt searches, respectively. Details on the pNN training and on the interpolation between mass points are given in Sections~\ref{sec:res_event_cat} and~\ref{sec:sig_bkg_modeling}. 

In each search, maximum likelihood fits are performed to the \mgg distributions of the corresponding event categories. A likelihood function is defined for each analysis category using analytic probability density functions (pdf) of \mgg for the signal and backgrounds. The smoothly falling background (continuum) is modeled directly from data. The signal and the single \PH production, which constitutes a resonant background, are modeled from simulation. In the \XYggHtt search, there is an additional background from Drell--Yan (DY) process where the two electrons from a \PZ decay are misidentified as photons. This background is modeled from data using an ABCD method~\cite{Choi:2019mip}. The fits are used to extract 95\% \CL upper limits on the relevant production cross sections.

\section{Event selection\label{sec:event_reco}}
Events are required to have at least two photon candidates. Photons are required to have $\abs{\eta} < 2.5$, excluding the barrel-endcap transition region ($1.442 < \abs{\eta} < 1.566$). Depending on their psudorapidity they are labelled as barrel photons ($\abs{\eta} < 1.442$) or endcap photons ($1.566 <\abs{\eta} < 2.5$). The variables used to select photon candidates include:
\begin{itemize}
    \item Charged-hadron isolation ($\mathcal{I}_{\text{ch}}$), the \pt sum of charged hadrons within a cone of $R=0.3$ centered around the photon direction.
    \item Photon isolation ($\mathcal{I}_{\text{ph}}$), the \pt sum of photons within a cone of $R=0.3$ centered around the photon direction.
    \item Track isolation ($\mathcal{I}_{\text{tk}}$), the \pt sum of all charged-particle tracks within a cone of $R=0.3$ centered around the photon direction.
    \item Hadronic over electromagnetic energy ratio (H/E), the ratio of the energy in the HCAL tower behind the photon supercluster to the energy of the photon's supercluster.
    \item \RNINE, the energy sum of the $3{\times}3$ crystals centered around the most energetic crystal in the cluster divided by the energy of the photon.
    \item $\sigma_{i\eta i\eta}$, the energy-weighted spread in $\eta$ of the $5{\times}5$ crystals centered around the most energetic crystal in the cluster~\cite{CMS:2020uim}.
\end{itemize}
The isolation variables are corrected to mitigate the contribution from pileup~\cite{CMS:2020uim}. Selected photons are required to have $\pt > 25\GeV$, $\mathcal{I}_{\text{ch}} < 20\GeV$, $\mathcal{I}_{\text{ch}} / \pt^\PGg < 0.3$, $\text{H/E} < 0.08$, and pass a ``conversion-safe'' electron veto criterion that rejects photons consistent with any charged particle tracks not matched to a conversion vertex~\cite{CMS:2015myp}. Further requirements are applied separately for barrel and endcap photons in the low \RNINE category. These requirements are shown in Table~\ref{tab:photonpreselcuts} for the respective \RNINE ranges. No further requirements are applied to the high \RNINE category ($>$0.85 and $>$0.90 for barrel and encaps, respectively).  Photon candidates are also required to satisfy a loose identification criterion based on a BDT classifier trained to separate photons from jets~\cite{CMS:2020uim}. The BDT is trained using variables that describe the shape of the photon's electromagnetic shower and isolation.

\begin{table}
    \centering
    \topcaption{Additional photon requirements for barrel and endcap photons at different ranges of \RNINE, intended to mimic the HLT requirements. \label{tab:photonpreselcuts}} 
    \begin{tabular}{ccccc}
                               & \RNINE       & $\sigma_{i\eta i\eta}$ & $\mathcal{I}_{\text{ph}}$ ({\GeVns}) & $\mathcal{I}_{\text{tk}}$ ({\GeVns}) \\
      \hline
      \multirow{1}{*}{Barrel}  & [0.50, 0.85] & $<$0.015               & $<$4.0                    &  $<$6.0                   \\
      \multirow{1}{*}{Endcaps} & [0.80, 0.90] & $<$0.035               & $<$4.0                    &  $<$6.0                   \\
    \end{tabular}
\end{table}

All events are required to have at least one diphoton candidate and in the case where several candidates exist, the candidate with the highest scalar sum \pt of the photons is chosen. Additionally, at least one photon in the pair must have $\pt > 30\GeV$.
Diphoton candidates are required to have an invariant mass (\mgg) in the range 100--180\GeV for the searches targeting processes with a \HGG~decay. For the \Ygg decay, this range is extended to 65--150\GeV and 100--1000\GeV for the low-mass and high-mass searches respectively. In all but the low-mass \XYggHtt search, we impose mass-dependent photon \pt requirements of $\pt/\mgg> 1/3$ and ${>}1/4$ for the highest \pt (leading) and second-highest \pt (subleading) photons, respectively.

When constructing an $\PH/\PY \to \PGt\PGt$ candidate, both hadronic and leptonic decay modes of the $\PGt$ lepton are considered. Therefore, three primary physics objects are considered: $\Pe$, $\PGm$, and $\tauh$.

Electron candidates are required to have $\pt > 10\GeV$ and $\abs{\eta} < 2.5$, excluding the barrel-endcap transition region ($1.442 < \abs{\eta} < 1.566$). Furthermore, the electron's impact parameter with respect to the PV in the transverse plane ($d_\mathrm{xy}$) is required to be less than 0.045\cm, and the distance between the vertex and electron along the beam axis ($d_\mathrm{z}$) is required to be less than 0.2\cm.

Muon candidates are selected by demanding: $\pt > 15\GeV$, $\abs{\eta} < 2.4$, $d_\mathrm{xy} < 0.045$\cm, $d_\mathrm{z} < 0.2$\cm. In addition, isolation and identification requirements are imposed on both electron and muon candidates~\cite{CMS:2015xaf,CMS:2018rym}. Finally, an isolation requirement of $\Delta R = \sqrt{\smash[b]{\Delta\eta\,^2 + \Delta\phi\,^2}} > 0.2$ with respect to selected photons is placed.

The $\tauh$ candidates are required to have $\pt > 20\GeV$, $\abs{\eta} < 2.3$, $d_\mathrm{z} < 0.2$\cm. Additionally, an isolation requirement of $\Delta R > 0.2$ with respect to the selected photons and leptons is placed. Finally, a selection is applied using the \textsc{DeepTau} discriminants. For genuine \tauh candidates with $\pt \in [30,70]\GeV$ from $\PH \to \PGt\PGt$ events, the selections result in identification efficiencies of 99\%, 99.95\%, and 80\% for $D_\Pe$, $D_\PGm$, and $D_\text{jet}$, respectively~\cite{CMS:2022prd}.

If only one \tauh candidate is found in the event, with no additional electrons or muons passing the selection, further tau lepton candidates are searched for as isolated tracks. These tracks must be associated with PF candidates, originate from the PV, have $\pt > 5\GeV$, $d_\mathrm{xy} < 0.2$\cm and $d_{z} < 0.1$~cm. An isolation requirement of $\Delta R > 0.2$ is applied with respect to the selected photons, electrons, muons, and \tauh.

A $\PGt\PGt$ candidate can be identified from any of the following pairs of reconstructed objects: $\tauh\tauh, \; \tauh\PGm, \; \tauh \Pe, \; \PGm \Pe, \; \PGm\PGm, \; \Pe\Pe, \; \tauh + \text{isolated track}$. Any event with an opposite-sign same flavor (OSSF) $\Pe\Pe$ or $\PGm\PGm$ pair is rejected if either: $m_{\ell\ell} \in [80,100]\GeV$ or $m_{\ell\ell\PGg} \in [86, 96]\GeV$, where $m_{\ell\ell\PGg}$ is calculated with respect to both the leading and subleading photons, and $m_Z$ is the nominal \PZ boson mass. This selection rejects events consistent with a \PZ $\to$ $\ell\ell$ or \PZ $\to$ $\ell\ell \PGg$ decay. Events are selected if they have a valid $\PGt\PGt$ candidate or if they have a single \tauh candidate.

In this analysis, information about jets in the event is used by the multivariate techniques described in Sections~\ref{sec:nonres_event_cat} and~\ref{sec:res_event_cat} to reject backgrounds, such as \ttbar production. Selected jets are required to have $\pt > 25\GeV$, $\abs{\eta} < 2.4$, and to be isolated from photons, leptons, or \tauh candidates, which is attained by requiring $\Delta R > 0.4$ between the jet and those objects.

\section{Nonresonant search event categorization\label{sec:nonres_event_cat}}
After the selection described in Section~\ref{sec:event_reco} is applied, events are categorized according to a discriminant designed to separate signal from background. In the nonresonant search, this discriminant is the output of a BDT trained with the \textsc{XGboost}~\cite{xgboost} framework on simulated samples of signal and background processes.
The nonresonant background simulated samples include \gjet, \dipho, \TTJETS, \ttgamma, \ttgammagamma, and \vgamma processes, while the resonant background simulated samples include \ggH, \VBF, \VH, and \ttH.
The samples used for training are independent from the ones used to evaluate the categorization performance and to create the final signal and resonant background models. This is also the case in the resonant searches.

Input features of the BDT include kinematical properties of the physics objects in each event: jets, leptons, photons, jet and lepton multiplicities, \ptmiss, and the \PQb tagging scores of jets from the \textsc{DeepJet} algorithm~\cite{CMS:2017wtu,Bols:2020bkb,CMS-DP-2018-058}. In channels that include more than one tau lepton candidate, a likelihood-based estimate of the invariant mass \mtt~\cite{Bianchini:2014vza} is also included. This estimation combines the measurement of \ptmiss and measurements of the visible $\PGt\PGt$ decay products. A further consideration in the input feature selection is the effect the BDT selection could have on the \mgg spectrum. The background modeling later described in Section~\ref{sec:bkg_modeling} requires the \mgg distribution of the continuum background to be smoothly falling such that the analytical functions chosen can model it well. It is important that no peaking structure is induced in the distribution which could lead to an artificial excess. Therefore, \mgg is not included as an input feature, and the \pt of the photon candidates are divided by \mgg to reduce correlations between the photon's \pt and \mgg. The \mgg distributions of the continuum background in simulation after the BDT selection are studied and no peaking structure is found.

\begin{figure}
    \centering
    \includegraphics[width=0.7\textwidth]{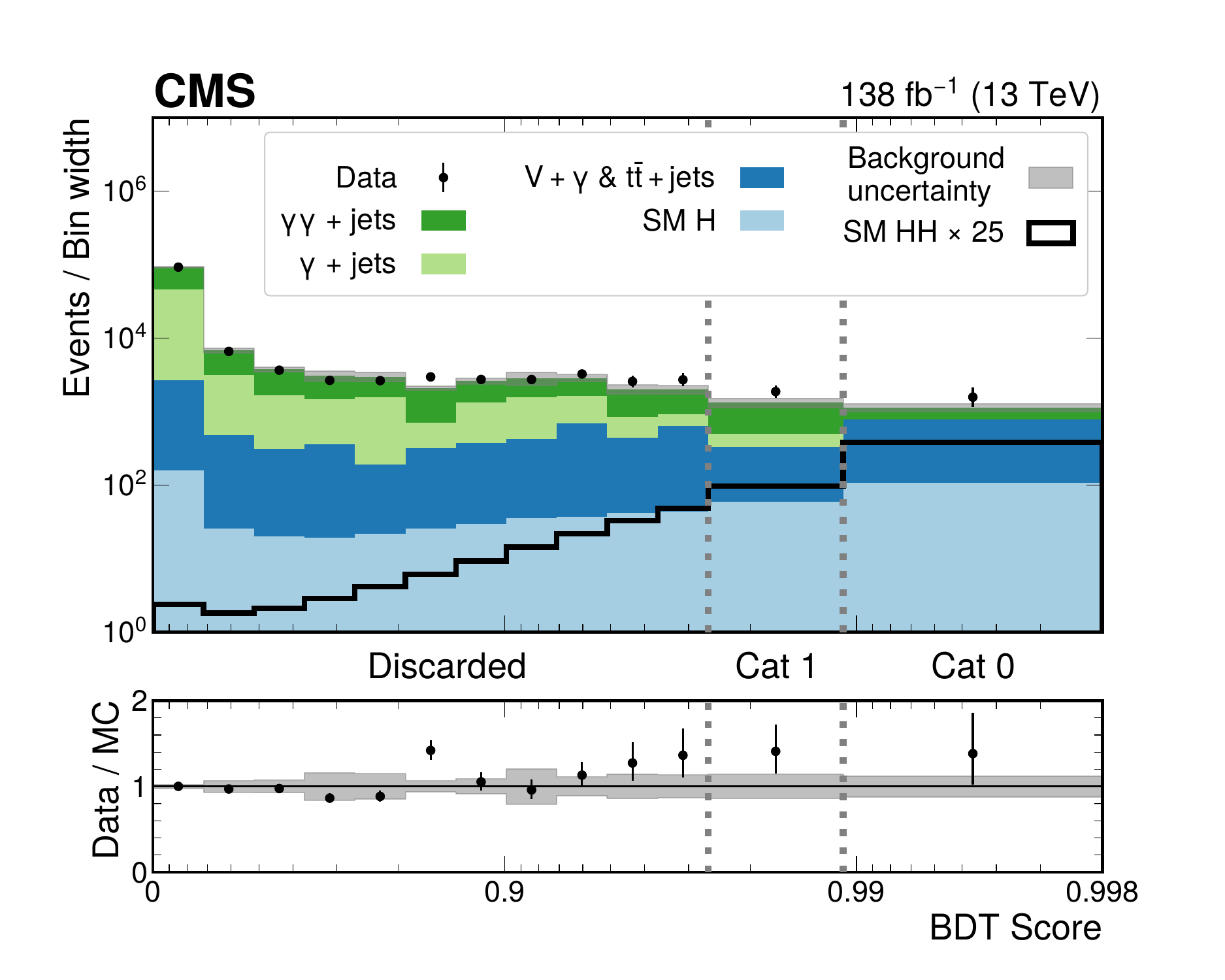}
    \caption{Distribution of the BDT scores used for the nonresonant analysis event categorization from data (black points) and predictions from MC simulation (colored histograms).
    The ``SM H'' process includes \ggH, \VBF, \VH, and \ttH, and the \TTJETS process also includes the \ttgamma and \ttgammagamma processes.
    The background histograms are stacked, while the signal distribution is shown separately. 
    The normalization of the signal distribution is set to 25 times the SM prediction and the background MC simulation is normalized to data.
    The ratio of the data to the sum of the background predictions is shown in the lower panel.
    Statistical MC uncertainties for the background are represented by the gray shaded bands.
    The gray dotted lines represent the boundaries that define the two analysis categories (Cat 0 and Cat 1). The rest of the data, labeled as `discarded' is not used in the fit to data.\label{fig:bdt_nonres}
    }
    
\end{figure}

The output score distribution from the BDT is shown in Fig.~\ref{fig:bdt_nonres}. Events are then categorized according to boundaries placed on the output score. The boundaries are set by optimizing the expected upper limit on the cross section of nonresonant \HH production, where the limit is calculated with signal and background yields from simulation and without systematic uncertainties included. Events are discarded if they have a BDT score ${<}0.974$. Two categories are defined with the rest of the events. The category purest in nonresonant \ggF \HH production (Cat 0), with a BDT score ${>}0.989$, has a signal efficiency with respect to the inclusive phase space of 10.5\%. The second category (Cat 1), with a BDT score ${<}0.989$, has a signal efficiency of 3.5\%. The possibility of having a greater number of categories was investigated but the improvement to the expected upper limit was found to be negligible.

\section{Resonant searches event categorization\label{sec:res_event_cat}}
In the resonant search, the discriminants used for event categorization are the outputs of pNNs which were first introduced in Section~\ref{sec:strategy}. Events passing the initial selection are used to train a pNN in each resonant search. The background simulation includes the \dipho, \ttgamma, \ttgammagamma, and \vgamma processes. The \gjet and \TTJETS processes are not used in training because they have limited sample sizes after the initial selection and lead to overtraining. 
 
Each signal sample (one for every mass hypothesis) is then weighted such that they all have an equal sum of weights. In each training batch, there are an equal number of signal and background events. These events are randomly sampled from the entire data set with probabilities given by their weights. Once a batch is created, the weight of every event is set to be unity so that the events are unweighted in the loss function.

A neural network that is trained to discriminate background from signal for a single mass hypothesis ($\mx=m_\text{i}$) has an output score that can be written as $f^i(\vec{x})$ where $\vec{x}$ is a vector of input features. In the case of a pNN, the function that we want to obtain is instead $f(\vec{x};\mx)$, where $f(\vec{x};\mx=m_i) = f^i(\vec{x})$. This way we can have a single network, which adapts the way it discriminates based upon the target mass hypothesis.

To achieve the intended parameterized behavior, the pNN is trained on simulated events from background and from all the signal hypotheses simultaneously. The parameters are included as additional input features, whose distributions are identical for the signal and background processes by construction. The simulated signal events for the \XHH (\XYH) processes are assigned values of \mx (and \my), which correspond to the simulation sample they originated from. For the sake of assigning \mx (and \my) to the background events, each background event in a batch is randomly paired with a signal event, and the \mx (and \my) values from the signal event are given to the background event. This ensures that on a batch-by-batch basis, there is no direct discrimination power provided by the parameters. Therefore, the network's dependence on the parameters is motivated purely by the correlation between the parameters and the rest of the variables. 

To check whether the pNN performs well for mass hypotheses not used in training, tests are performed where the pNN is trained using samples from all nominal hypotheses except one, and the performance of that pNN for the excluded hypothesis is compared to the pNN which is trained on all hypotheses. This test is performed in every search and for all mass hypotheses which are not at the boundaries of the mass ranges. At a background efficiency of 1\%, the relative changes to the signal efficiency are less than 1\% for all but three of the tests, and less than 2\% for all of them. 

The input features for the resonant pNNs are the same as those used for the nonresonant search BDT with the addition of the invariant mass of the \ggtt system, $m_{\ggtt}$, which includes a \ptmiss correction. As in the nonresonant search, the continuum background distributions of \mgg are studied in simulation to look for peaking structures induced by the pNN selection. In the high-mass \XYggHtt search, which has a significantly different range in \mgg compared to the other searches, the exclusion of $m_{\ggtt}$ and $m_{\ell\ell\PGg}$ from the input features was found necessary to prevent sculpting. In all other searches, no peaking structures were found.

For each mass hypothesis, we perform a probability integral transform so that the distribution of the output score for the simulated background events is uniform. This transformation is applied to both data and simulation before defining the pNN boundaries for each category. This choice helps with the interpolation of the signal efficiency because a set of boundaries on a pNN score now corresponds to the same efficiency on the simulated background events, regardless of \mx (and \my). Given also that the network input distributions of the signal evolve in a continuous way with \mx (and \my), it is expected that the signal efficiencies for a given background efficiency, or equivalently, a set of pNN boundaries, will evolve in a predictable manner. In Section~\ref{sec:sig_bkg_modeling}, one finds that this is indeed the case.

\begin{figure}
    \centering
    \includegraphics[width=.49\textwidth]{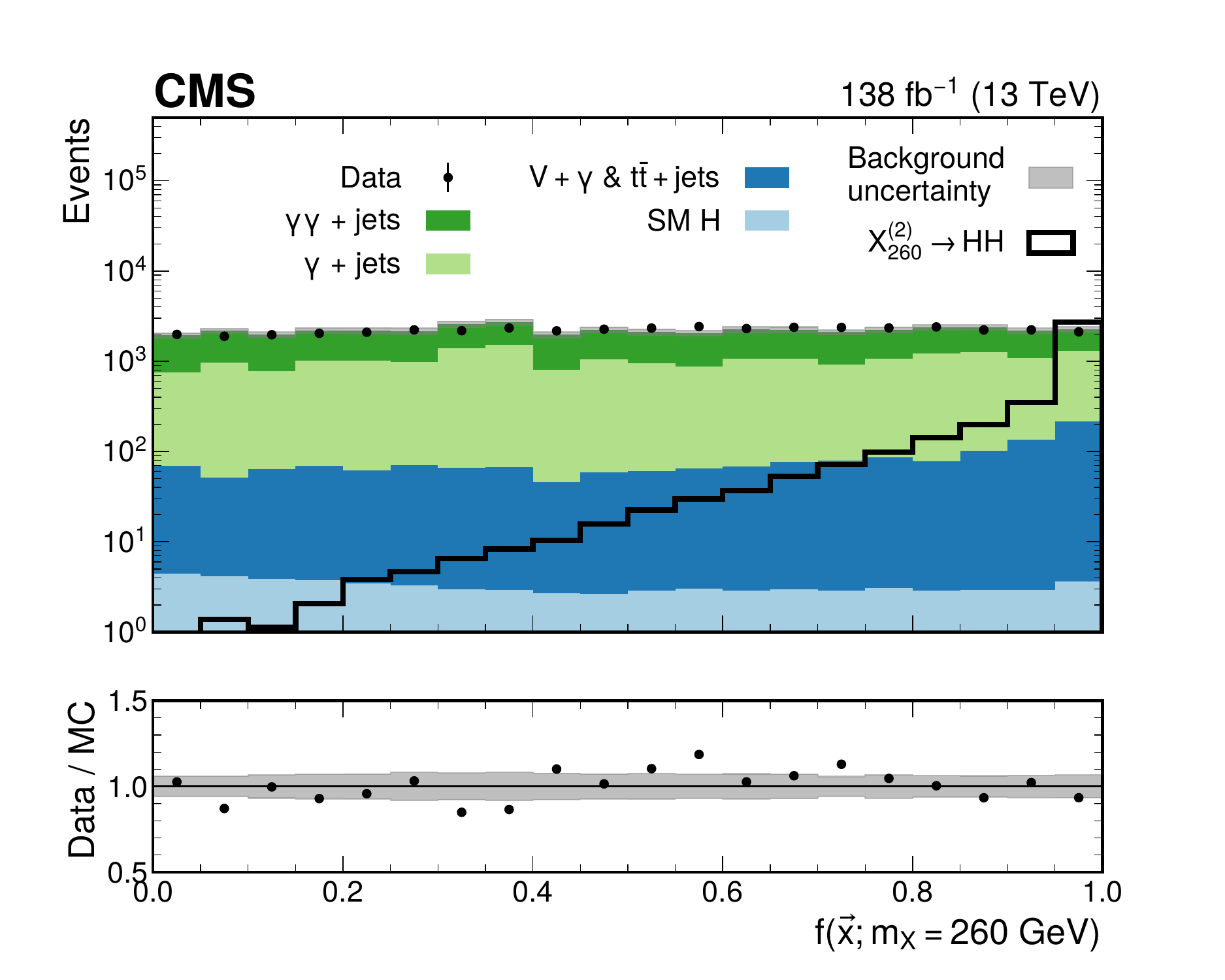}
    \includegraphics[width=.49\textwidth]{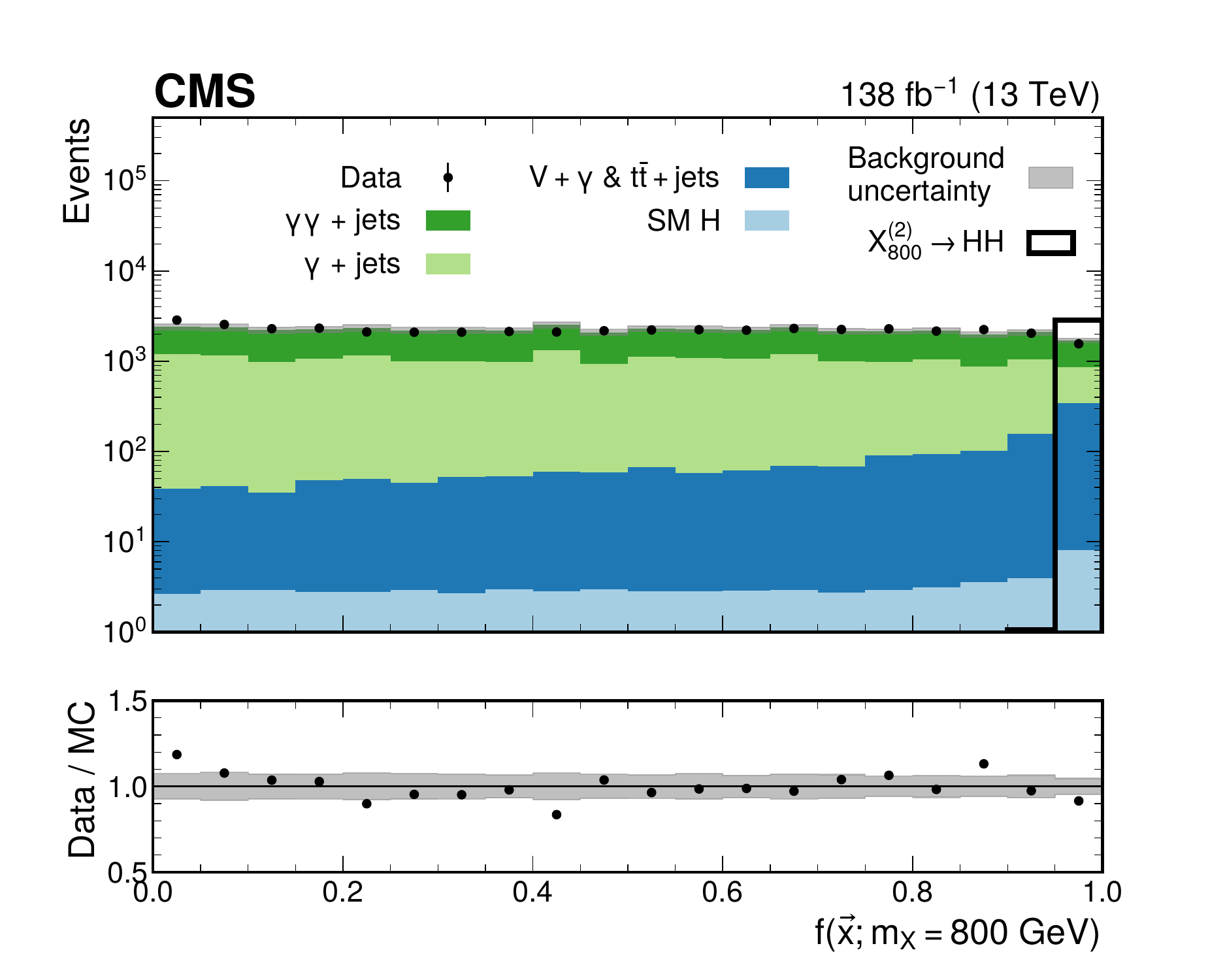}
    \caption{Transformed output of the pNN used in the \XTwoHH search, evaluated at $\mx=260\GeV$ (left) and 800\GeV (right). 
    The filled histograms represent the background simulation, and the data are shown by the black points. 
    The ``SM H'' process includes \ggH, \VBF, \VH, and \ttH, and the \TTJETS process also includes the \ttgamma and \ttgammagamma processes. 
    The targeted signal distributions for which the pNN is evaluated are shown by the black unfilled histograms. 
    The background MC simulation is normalized to data and the signal is normalized to an arbitrary cross section for representation purposes. 
    The ratio of the data to the sum of the background predictions is shown in the lower panel. 
    Statistical MC uncertainties for the background are represented by the gray shaded bands.\label{fig:pnn_xhh_evolution}}
\end{figure}

The transformed pNN output score distributions evaluated at $\mx=260\GeV$ and 800\GeV for data and simulated signal and background events (including \gjet and \TTJETS) are shown in Fig.~\ref{fig:pnn_xhh_evolution} for the \XTwoHH search. The plots demonstrate how the discriminating power changes as a function of \mx. At the high \mx value of 800\GeV, the signal distribution has migrated towards higher scores, showing it is easier to discriminate a high \mx signal from background. In addition, the background composition in the signal-like bins changes. For example, the relative contribution to the background from single \PH production is larger for $\mx=800\GeV$, compared to $\mx=260\GeV$.

\begin{figure}
    \centering
    \includegraphics[width=.49\textwidth]{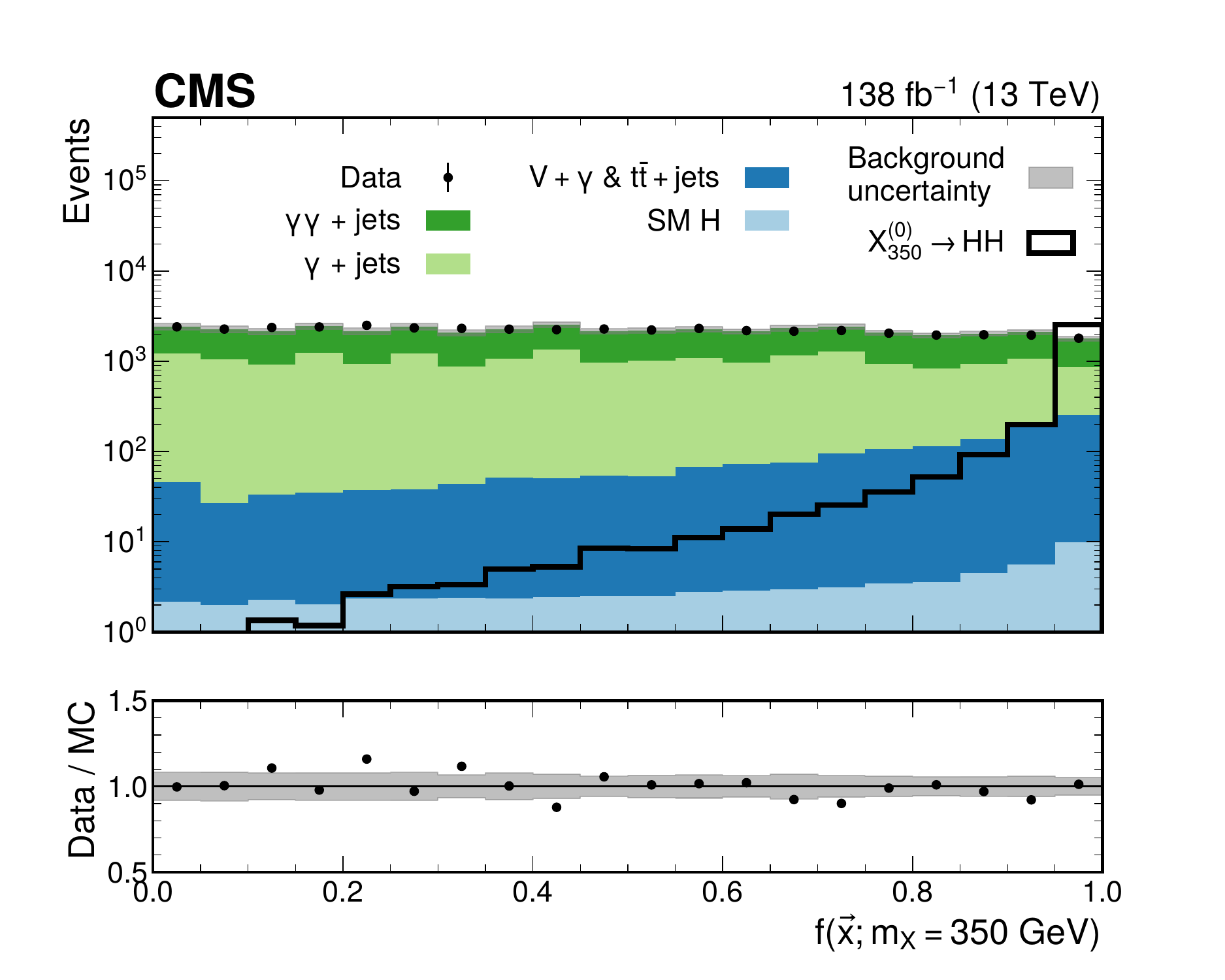}
    \includegraphics[width=.49\textwidth]{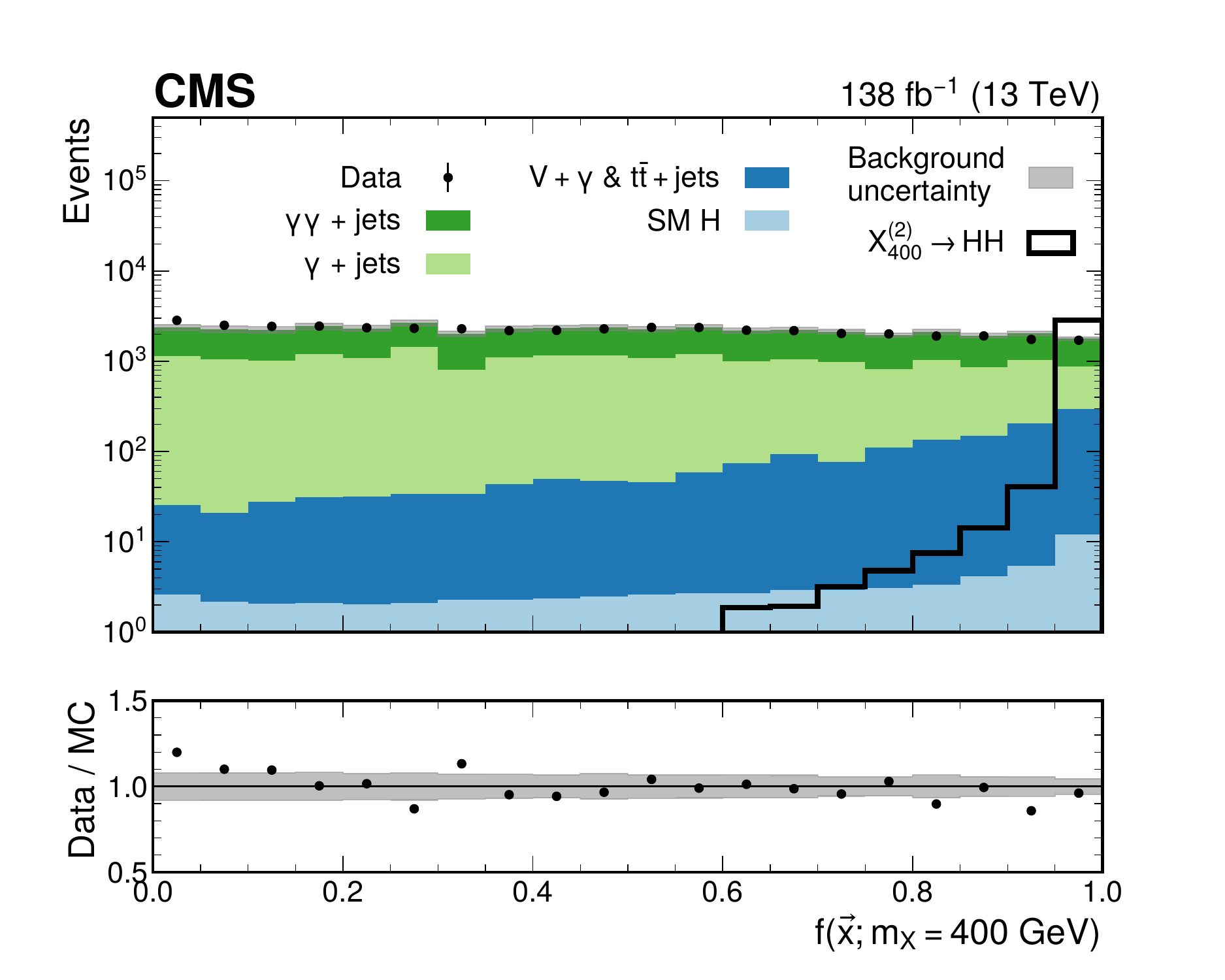}
    
    \includegraphics[width=.49\textwidth]{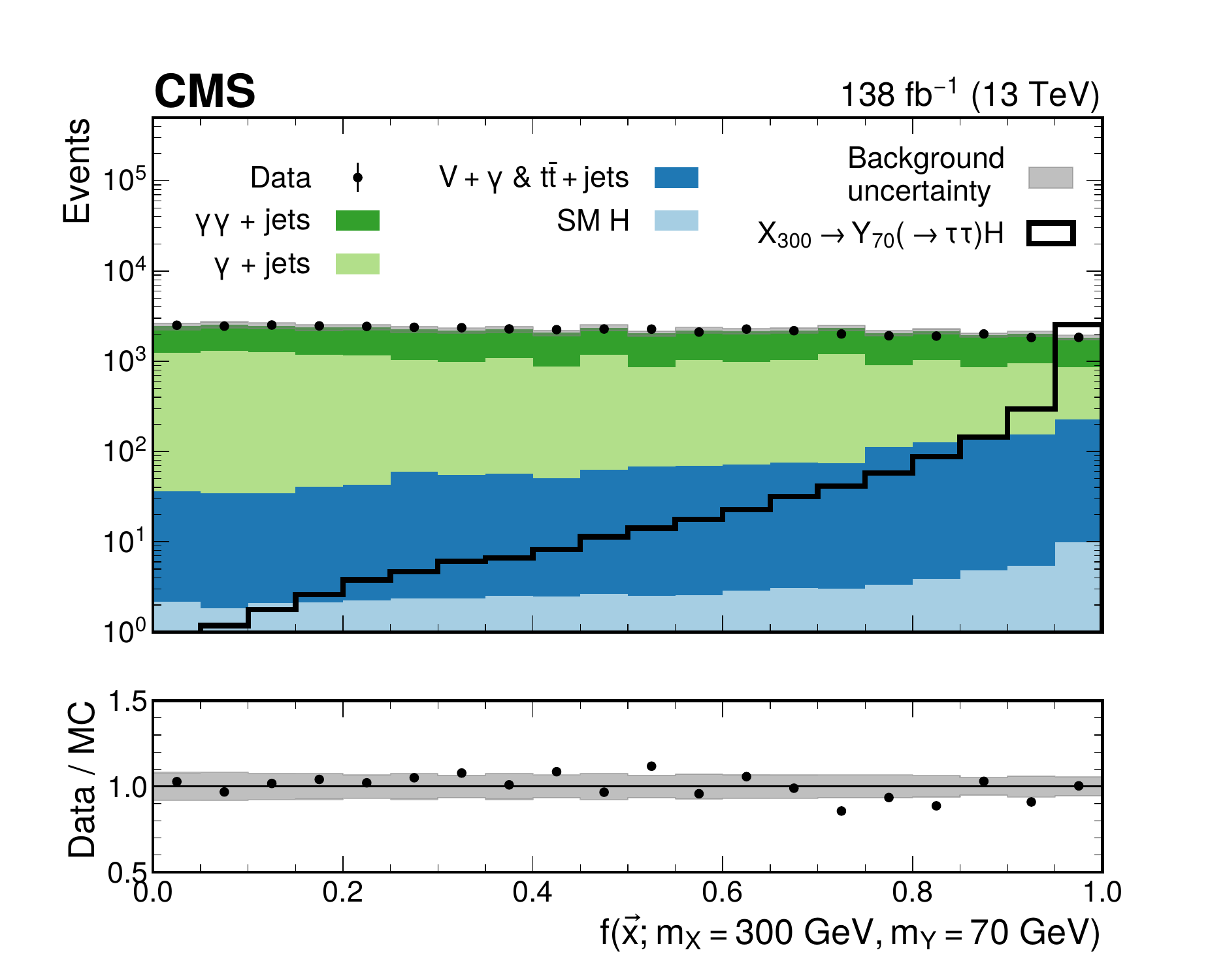}
    \\
    \includegraphics[width=.49\textwidth]{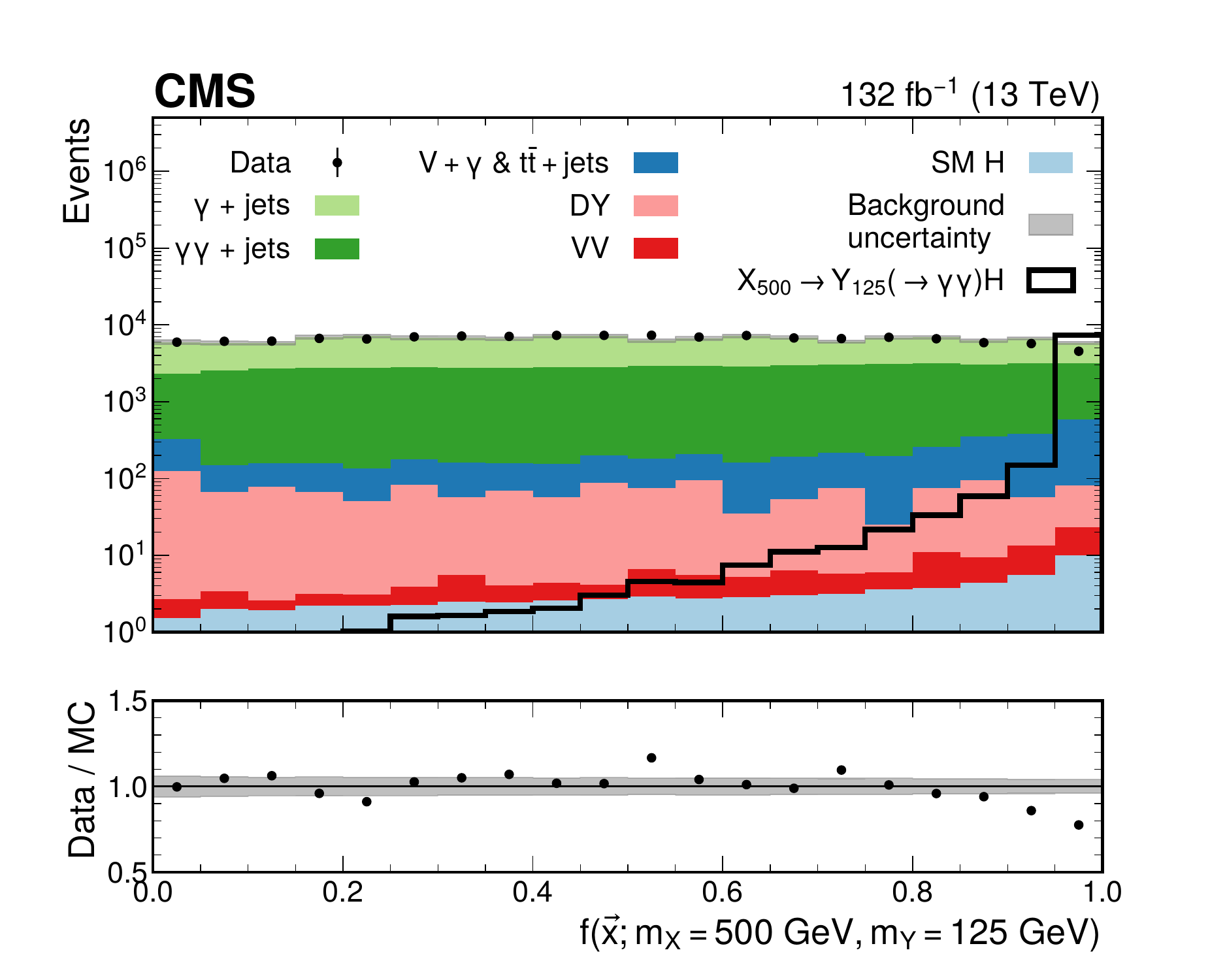}
    \includegraphics[width=.49\textwidth]{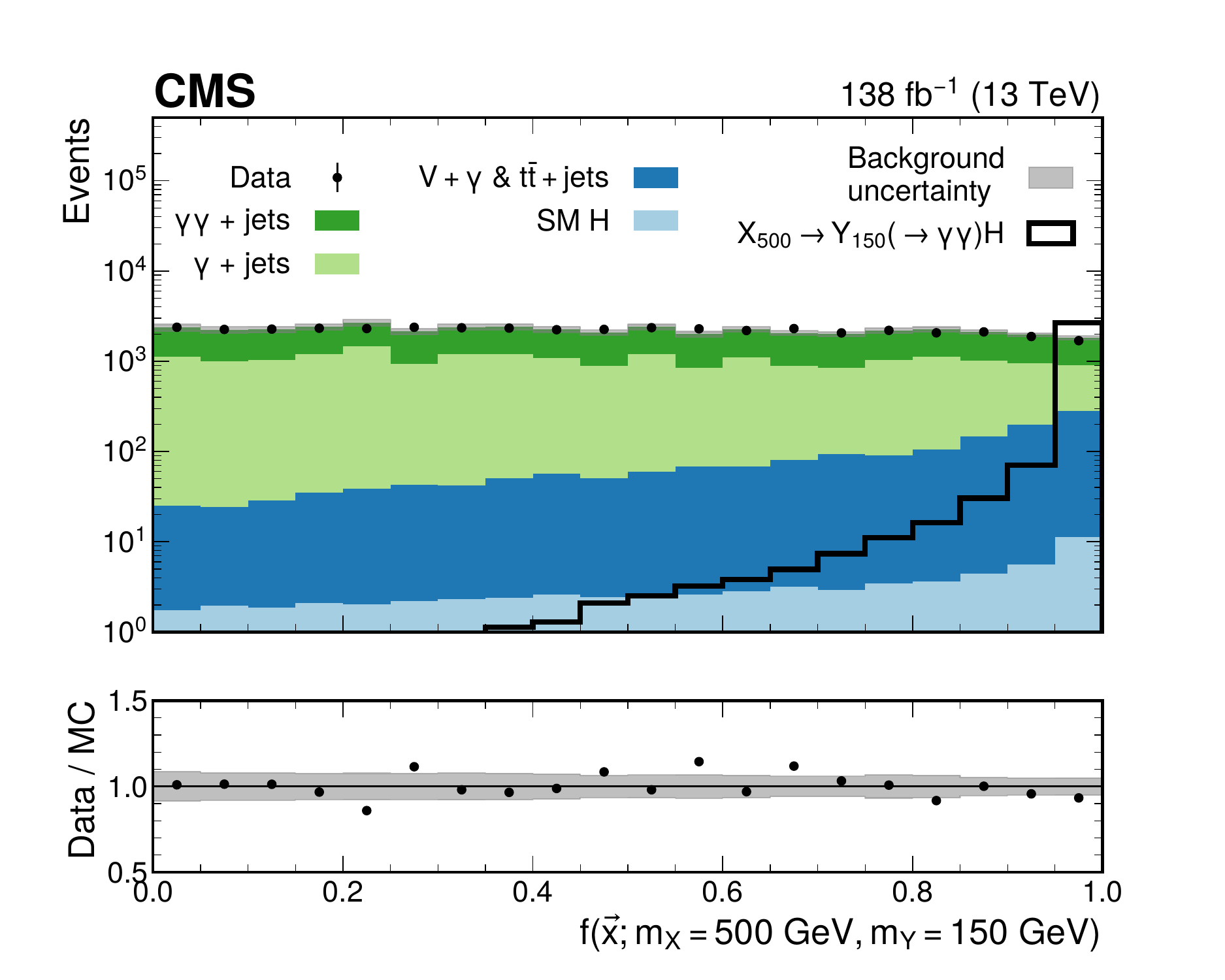}
    \caption{Transformed output of the pNNs used in the \XZeroHH (upper left), \XTwoHH (upper right), \XYttHgg (middle), low-mass \XYggHtt (lower left) and high-mass \XYggHtt (lower right) searches.
    The pNNs are evaluated at the mass points where the largest excess with respect to the background-only hypothesis is observed. 
    If the MC simulation at this mass point is not available, then the sample produced at a mass point closest to the excess is shown.
    The filled histograms represent the background simulation, and the data are shown by the black points. 
    The ``SM H'' process includes \ggH, \VBF, \VH, and \ttH, and the \TTJETS process also includes the \ttgamma and \ttgammagamma processes.
    The targeted signal distributions for which the pNN is evaluated are shown by the black open histograms.
    The background MC simulation is normalized to data and the signal is normalized to an arbitrary cross section for representation purposes. 
    The ratio of the data to the sum of the background predictions is shown in the lower panel. 
    Statistical MC uncertainties for the background are represented by the gray shaded bands.\label{fig:pnn_resonant}}
\end{figure}

Examples of transformed pNN output score distributions for all resonant searches are shown in Fig.~\ref{fig:pnn_resonant}. For each search, taking into account the search results discussed in Section~\ref{sec:results}, the pNN is evaluated at the mass hypothesis where the largest excess with respect to the background-only hypothesis is found in data.

Event categories are defined by placing boundaries on the transformed pNN output scores according to the expected number of simulated background events (including \GAMJETS and \TTJETS) in each category. The events are ordered by their pNN scores and categorized starting from the highest pNN scores. A minimum number of expected events in a category was found needed to reduce bias originating from the choice of function used to model the background \mgg distribution. This bias is reduced by using the discrete profiling method~\cite{Dauncey:2014xga} but is not completely eliminated. To quantify the residual bias, a large number of pseudo-datasets are generated for a representative set of mass points and significances are calculated for each pseudo-dataset when fixing the background function to the one used to generate the datasets and when allowing the background function to float. For a given mass point, the fraction of datasets with a significance higher than $X$, where the range $0 < X < 3$ is considered, were compared and the maximum difference was taken to be the residual bias. A minimum number of 10 expected events was chosen since it was the lowest number (and therefore most sensitive) which still led to acceptable bias of 20\%.

In each resonant search, event categories are sequentially defined, starting with 10 expected background events, and continuing to add categories with 10 events, evaluating the expected limits on $\sigma(\XYHHggtt)$ with every category added. This continues until none of the limits improve by more than 1\%, at which point a category with twice the number of events is considered instead. This continues until the sum of events across all categories is greater than the number in data after preselection. 
This results in four categories of 10 events, one category of 20 events and one category of 80 events for each search, going from the most to least sensitive, and the \XYH searches have a final, extra category of 320 events. 

In the \XTwoHH search, the expected limits at the nominal mass hypotheses using this categorization were compared to those found using an alternative categorization strategy that was performed for each hypothesis individually, and was a grid search of category boundaries aimed at minimizing the expected limit. For $\mx > 320\GeV$, the grid-search approach improved the limits by at most 0.2\%, and for $\mx < 320\GeV$, by at most 1.4\%. The categorization strategy based upon expected background events is used despite this because it can be applied to the intermediate mass hypothesis too, which is not possible with the grid-search approach, and the losses in sensitivity at the nominal mass hypotheses are considered small.

\section{Signal and background modeling\label{sec:sig_bkg_modeling}}
This section describes the construction of the signal and background models used in the maximum likelihood fits. For the nonresonant \HH, \XZeroHH, \XTwoHH and \XYttHgg searches, the probability density functions (pdfs) are defined over the range of $100<\mgg<180\GeV$. In the high-mass \XYggHtt search, the pdfs are defined in the range of $100<\mgg<1000\GeV$, whilst in the low-mass \XYggHtt search, the pdfs are defined in the range of $65<\mgg<150\GeV$. These ranges ensure there is a sufficient number of background events that populate the \mgg distribution in order to accurately model the continuum background component. The lower bound at 65\GeV in the low-mass \XYggHtt search is chosen to avoid the trigger efficiency turn-on curve at low \mgg.

The shape and normalization of the signal models are derived assuming $\mH=125.38\GeV$, which is the most precise measurement of the Higgs boson mass reported by CMS~\cite{CMS:2020xrn}.

\subsection{Signal modeling in the nonresonant search}\label{sec:sig_model_nonres}
The shape of the \mgg distribution for signal events is derived using simulated events. This is done independently in each nonresonant search analysis category, for each year of data-taking. The signal pdf is constructed by fitting the \mgg distribution with a double Crystal Ball (DCB) function~\cite{Oreglia:1980cs,Gaiser:1982yw}.

\begin{figure}
    \centering
    \includegraphics[width=0.5\textwidth]{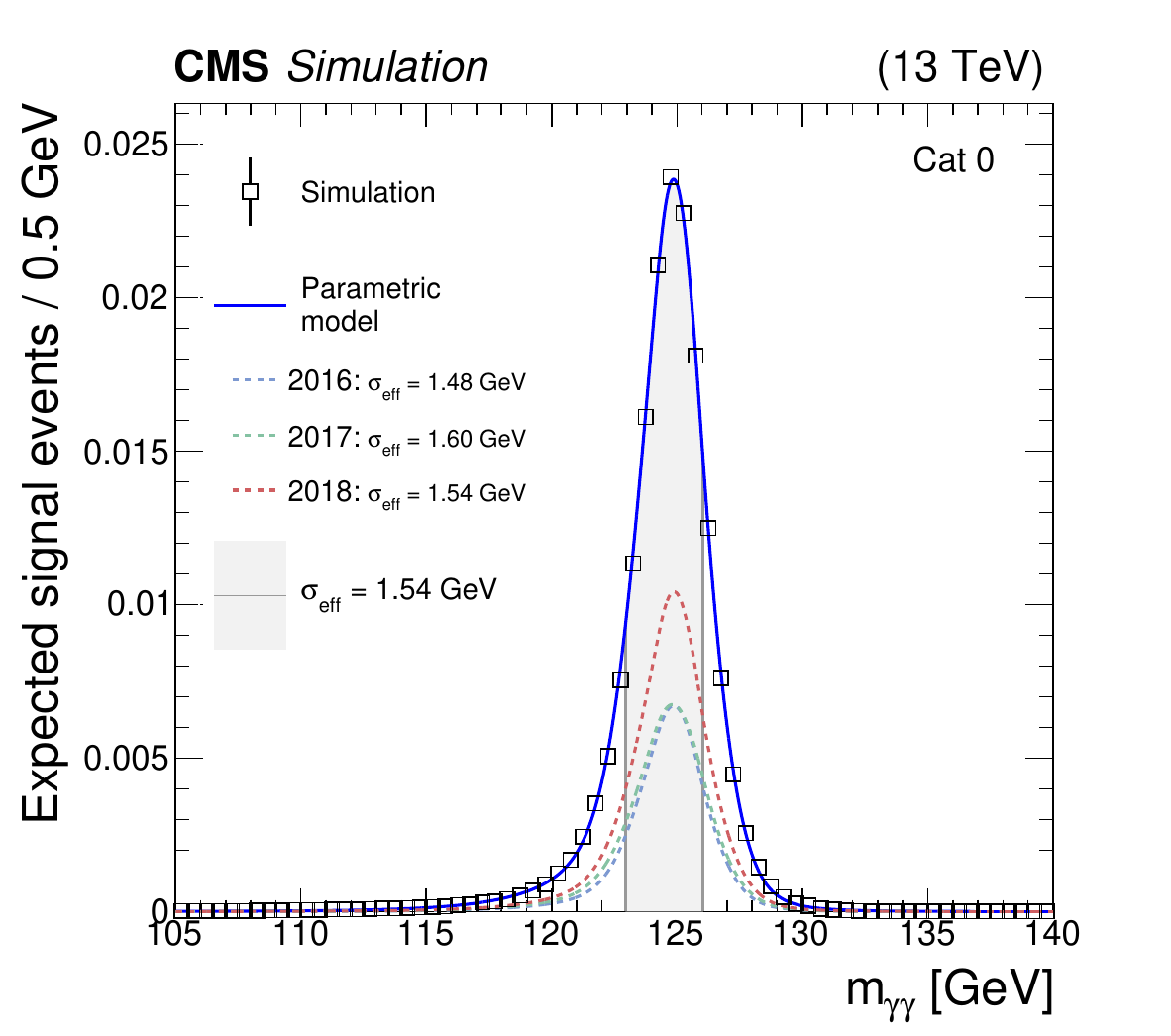}
    \caption{Signal model for the purest analysis category in the nonresonant search, shown for each year of simulated data, and for the sum of all years together. The open squares represent the weighted simulated events, whose uncertainty are smaller than the marker size, and the blue line is the corresponding model. The gray shaded areas correspond to the $\sigma_{\text{eff}}$. The contribution from each year of data taking is illustrated with the dotted lines. \label{fig:smodel_nonres}}
\end{figure}

The normalization of the signal model in analysis category, $i$, is defined by the formula,
\begin{equation}
    N_i = 2 \, \sigma({\Pp\Pp} \to \HH) \, \mathcal{B}(\PH \to \PGg\PGg) \, \mathcal{B}(\PH \to \PGt\PGt) \, \epsilon_i \,  \mathcal{L} \, \mathcal{C}(\vec{\theta}),
    \label{eq:norm_nonres}
\end{equation}
where $\sigma({\Pp\Pp} \to \HH)$ is the nonresonant \ggF HH cross section, on which an upper limit is placed, $\mathcal{B}(\PH \to \PGg\PGg)$ and $\mathcal{B}(\PH \to \PGt\PGt)$ are the SM branching fractions for \HGG and \Htt respectively, $\epsilon_i$ is the efficiency for signal events to be reconstructed in analysis category $i$, which is derived directly from simulation, and $\mathcal{L}$ is the integrated luminosity. The final term, $\mathcal{C}(\vec{\theta})$, corresponds to the corrections to the signal yield estimates from nuisance parameters, $\vec{\theta}$, representing systematic uncertainties in the analysis (see Section~\ref{sec:systematics}). Figure~\ref{fig:smodel_nonres} shows the signal model for the purest analysis category in the nonresonant search, which fits well to the simulated events and has a shape that is indicative of the rest of the signal models used in this analysis, including the resonant searches. The gray shaded areas correspond to the $\sigma_{\text{eff}}$, defined as half the width of the narrowest interval containing 68.3\% of the \mgg distribution. For the purest and second purest categories, $\sigma_{\text{eff}}$ is 1.54\GeV and 1.64\GeV respectively, and the shapes of the models are otherwise similar.

\subsection{Signal modeling in the resonant searches}\label{sec:sig_model_res}
Extracting limits for mass hypotheses, for which a simulation was not generated, introduces additional complexity to the signal modeling for the resonant search. The signal models for these hypotheses, denoted as intermediate mass points, are derived by interpolating both the shape and normalization from the neighbouring mass points in which simulated samples were generated, denoted as nominal mass points. For the \XZeroHH and \XTwoHH searches, the interpolation of parameters is performed in one dimension, \mx, whilst the \XYH searches require interpolation in two dimensions, \mx and \my.

As in the nonresonant search, DCB functions are used to model the signal. For the nominal mass points, the DCB shape parameters are extracted in a $\chi^2$ fit to the simulated data points. For the intermediate mass points, the mean, $\overline{m}_{\PGg\PGg},$ and width, $\sigma$, of the Gaussian core of the DCB are determined by interpolating the values at the nominal mass points with a linear spline. The rest of the shape parameters do not change significantly as a function of \mx and \my and therefore the values are simply taken from the closest nominal mass point. 

In the \XZeroHH, \XTwoHH and \XYttHgg searches, a set of analysis categories are defined for each considered mass hypothesis \mx (and \my). Signal events produced under the targeted \mx (and \my) hypothesis in analysis category, $i$, are normalized according to
\begin{equation}
    N_i(\mx,\my) = \sigma({\Pp\Pp} \to \PX \to (\PH/\PY)\PH \to \ggtt) \, \epsilon_i(\mx,\my) \, \mathcal{L} \, \mathcal{C}(\vec{\theta}).
\end{equation}
where the quantity $\sigma({\Pp\Pp} \to \PX \to (\PH/\PY)\PH \to \ggtt)$ is the BSM cross section on which an upper limit is placed. The key difference with respect to Eq.~\ref{eq:norm_nonres} is that the signal efficiency, $\epsilon_i$, becomes a function of \mx for the \XHH searches, or \mx and \my for the \XYttHgg search. 

\begin{figure}
    \centering
    \includegraphics[width=.49\textwidth]{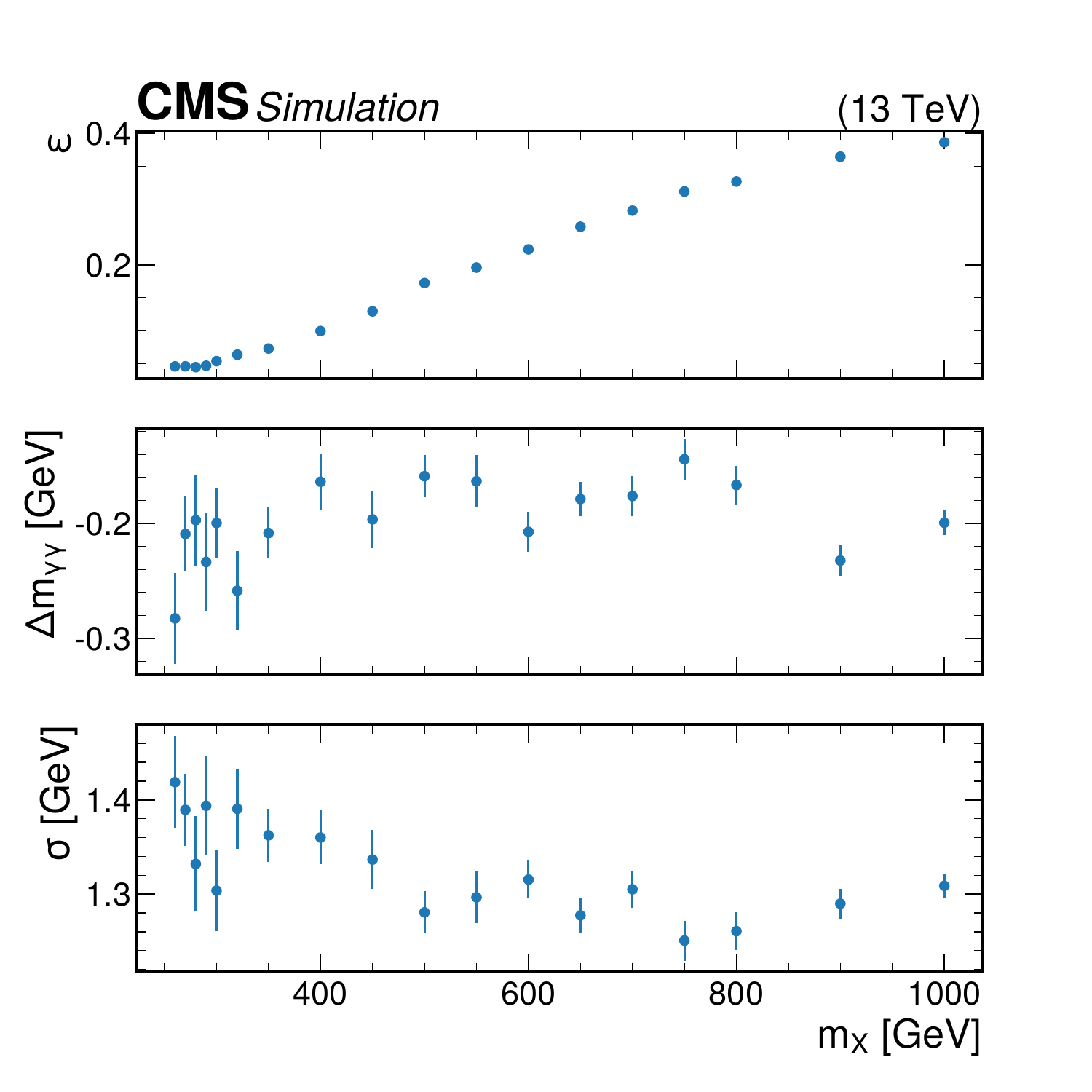}
    \includegraphics[width=.49\textwidth]{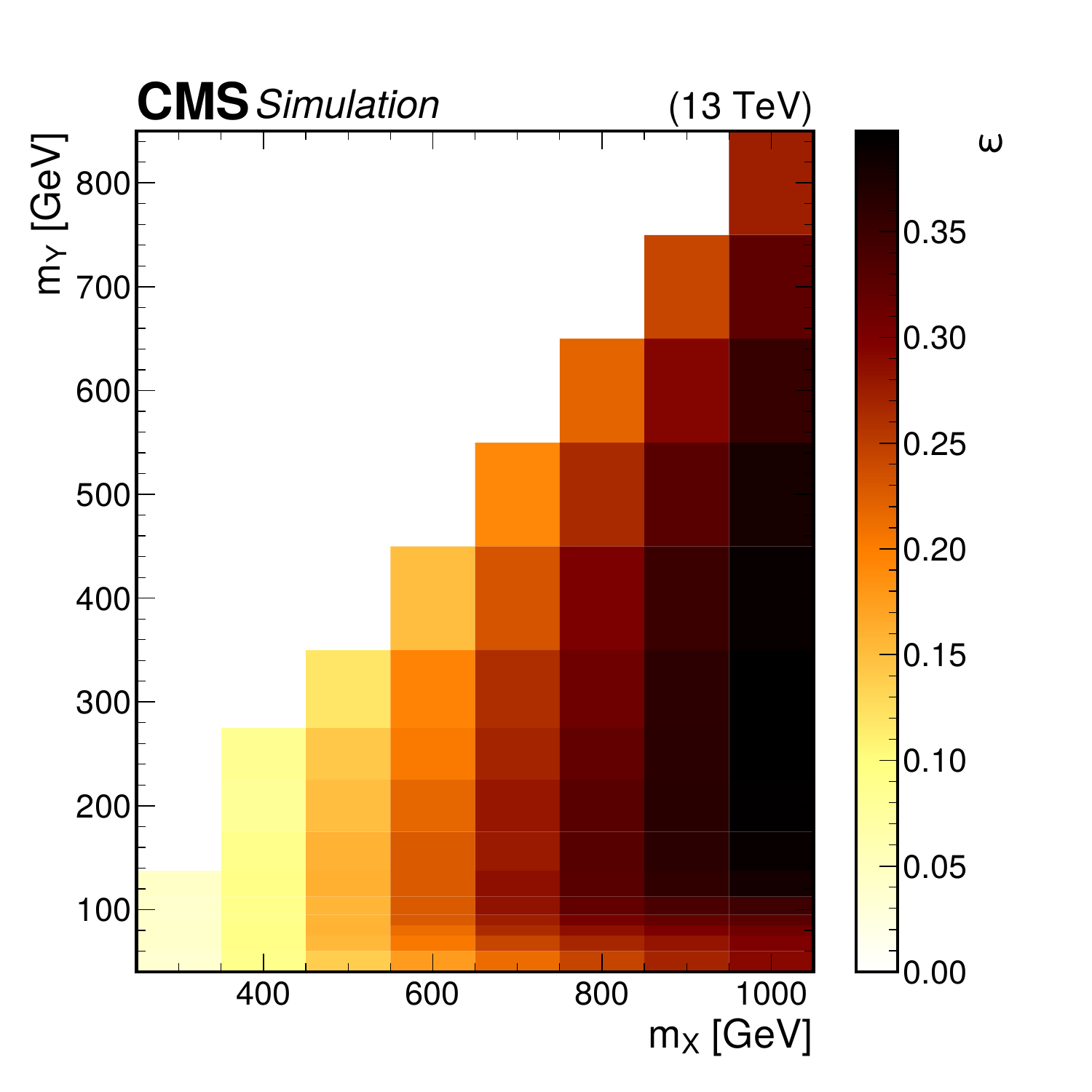}
    \caption{Signal efficiency, $\epsilon$, and interpolated DCB shape parameters, $\Delta m_{\PGg\PGg}$ and $\sigma$, for the highest purity analysis category in the \XTwoHH search, as functions of \mx (left).
    The first shape parameter, $\Delta m_{\PGg\PGg}$, is defined as $\overline{m}_{\PGg\PGg}-\mH$.
    Signal efficiency in the (\mx, \my) plane for the highest purity analysis category in the \XYttHgg search (right).\label{fig:smodel_res_spline}}    
\end{figure}

The signal efficiencies $\epsilon_i$ for the nominal mass points are computed directly from the simulation as the fraction of signal events that pass the selection for analysis category $i$. The signal efficiencies for the intermediate mass points are determined by interpolating the nominal mass point values using splines. Both linear and cubic splines are considered, where the type of spline used at a given mass point is the one that leads to the lowest interpolation systematic uncertainty (as discussed in Section~\ref{sec:systematics}). These splines are implemented using the \textsc{SciPy} package~\cite{2020SciPy-NMeth} where the cubic spline in 2D uses techniques described in Ref.~\cite{ALFELD1984169}. In Fig.~\ref{fig:smodel_res_spline} (left), the DCB shape parameters and signal efficiency for the highest purity analysis category are shown as functions of \mx, for the \XTwoHH search. Figure~\ref{fig:smodel_res_spline} (right) shows the signal efficiency in the (\mx, \my) plane, for the highest purity analysis category in the \XYttHgg search.

In the \XYggHtt search, the mean of the DCB's Gaussian core is instead parameterized by $\Delta m_{\PGg\PGg}= \overline{m}_{\PGg\PGg}-\my$ to account for the changing peak position as a function of \my. Additionally, the spacing in \my points in which a limit is calculated is driven by the experimental \mgg resolution ($\approx$1.5\GeV at $\my = 125\GeV$). Given the large \my range considered in this search, it is not feasible to define optimized analysis categories at every \my point. It is also unnecessary since the kinematical distributions of the signal, which the pNN mainly uses to discriminate with, do not change significantly over the scale dictated by the resolution. Therefore, analysis categories are constructed for a less granular set of (\mx, \my) hypotheses, and when placing a limit at a particular \my, the closest set of analysis categories is used. This feature introduces an additional complication to the signal model, as it requires interpolating the signal model parameters for nearby values of \my, for which the analysis category is not optimized. These secondary interpolations are derived for the $\epsilon$, $\Delta m_{\PGg\PGg}$, and $\sigma$ parameters by propagating the neighboring nominal mass point simulated samples into the analysis categories optimized for \my. Compared to the primary interpolation described above, the secondary interpolation has a smaller effect in general since the changes in \my that it interpolates over are smaller. Averaging over all mass points and categories, the changes to $\epsilon$, $\Delta m_{\PGg\PGg}$ and $\sigma$ at the maximal change in \my are 0.2\%, $0.3\GeV$, and $0.8\GeV$, respectively.

\subsection{Background modeling\label{sec:bkg_modeling}}
Background events consist of two contributions: the continuum background, which forms a smoothly falling distribution in \mgg, and events from single \PH production in the diphoton final state \HGG, which form a peak around \mH. The \HGG~contribution is modeled using the same procedure used for signal events and is normalized according to the best available cross section and branching fraction recommendations from the LHC Higgs Working Group~\cite{LHCHiggsCrossSectionWorkingGroup:2016ypw}, within appropriate theoretical uncertainties. The remainder of this section details the modeling of the continuum background component. In the final maximum likelihood fit, the total background model is taken as the sum of the continuum and \HGG~components.

The continuum background model is extracted directly from data using the discrete profiling method~\cite{Dauncey:2014xga}. With this technique, the systematic uncertainty associated with choosing a particular analytic function to fit the \mgg distribution is accounted. This works by introducing a discrete nuisance parameter into the maximum likelihood fit, which describes the choice of background function in an analysis category.

A large set of candidate function families is considered, including exponential functions, Bernstein polynomials, Laurent series, and power law functions. For each family of functions, an $\mathcal{F}$-test~\cite{10.2307/2340521} is performed to determine the maximum order of parameters to be used, whilst the minimum order is determined by placing a requirement on the goodness-of-fit to data. In the $\mathcal{F}$-test, each function is fit to the \mgg distribution by minimizing twice the negative logarithm of the likelihood (\TwoNLL). For the nonresonant \HH, \XZeroHH, \XTwoHH, and \XYttHgg searches, the fits are performed only in the mass sidebands, defined as ${\mgg \in [100,115] \cup [135,180]}\GeV$, to avoid introducing a bias from the signal region.

The fitting strategy for the \XYggHtt searches is slightly different, as it involves searching for a resonance \PY over a large range in \mgg. To account for this, the mass sidebands used to construct the continuum background model are defined by a sliding blinded region in \mgg,
\begin{equation}
    \my \pm 10\GeV  \frac{\my}{125.38\GeV},
\end{equation}
where the size of the blinded region increases as a function of \my to account for the increase in the resonance width. This formula holds for all considered \my, with the exception of $70<\my<72\GeV$ in the low-mass \XYggHtt search. For these \my points, data with $65<\mgg<68\GeV$ is kept to form a low-mass sideband; the fraction of signal events in this region from an $\my=70\GeV$ resonance is around 4\%, and therefore will have a negligible impact on the construction of the background model.

In the resonant searches, the Bernstein polynomials were removed from the envelope and replaced by an exponential of a polynomial function, which was found to provide a better description of the data over the invariant mass range used in these searches. For the same reason, in the high-mass \XYggHtt search, the so-called dijet functions~\cite{CMS:2022eud} are also added to the continuum background model.

When extracting the results in the final maximum likelihood fits, the discrete profiling method minimizes the overall \TwoNLL considering all allowed functions for each analysis category. Both the normalization and the shape parameters of these functions are included as free parameters in the fit. A penalty term equal to the number of parameters of the function is added to \TwoNLL to take into account the number of floating parameters in each candidate function, thus penalizing functions with high complexity. Checks are performed to ensure that modeling the continuum background \mgg distribution using the discrete profiling method introduces negligible bias to the final results.

\subsection{Drell--Yan modeling}
In the low-mass \XYggHtt search, there is an additional background from DY where two electrons from a \PZ boson decay can be misidentified as photons, leading to a peak in the \mgg spectrum close to the \PZ boson mass. To model this background contribution, an ``ABCD method"~\cite{Choi:2019mip} based on control samples in data is employed. Three control regions are defined by inverting the electron veto (region B), or inverting the pNN selection (region C) or inverting both (region A). This leaves the signal region, labeled as D. By inverting the electron veto, regions A and B are enriched in DY. These regions constrain the shape of the DY distribution in $\mgg$. In regions A and C, the yields for DY are extracted, constraining the electron veto efficiency.

Since kinematical variables are input to the pNN, and the shape of the diphoton mass peak is dependent mainly on the kinematical variables of the two photons, the shape of the DY process is assumed to be the same in regions with the same pNN selection (in A and C, and in B and D). For regions A and C, the shape is modeled by a Gaussian function, except in cases where a poor goodness-of-fit is found and a DCB is used instead. When extracting the final results, the shape parameters are allowed to float. Similarly, the rates of the DY background in regions A, B and C are allowed to float. The rate of DY in region D is given by the rate in region B, multiplied by the electron veto efficiency (rate in region C divided by rate in region A).

Finally, given that there are 7 pNN regions, this method is applied 7 times, where there are 7 independent B and D regions, and the A and C regions remain the same.

\section{Systematic uncertainties\label{sec:systematics}}
As described in Section~\ref{sec:sig_bkg_modeling}, the systematic uncertainty associated with the nonresonant background estimation is calculated with the discrete profiling method.
Systematic uncertainties affecting the signal models and/or the resonant background models are implemented in one of two ways.
First, uncertainties that modify the shape of the \mgg distribution are incorporated as nuisance parameters affecting the mean and width of the DCB functions that comprise the signal model.
Second, uncertainties that do not affect the shape of the \mgg distribution are implemented as log-normal uncertainties in the event yield.
Uncertainties of the first type are typically experimental sources of uncertainty that impact the energy of individual photons, while uncertainties of the second type include both theoretical and experimental sources, which have a negligible impact on the shape of the \mgg distribution.

\subsection{Theoretical uncertainties}
Sources of theoretical uncertainties considered in this analysis are those that affect the $\HGG$ branching fraction, the uncertainty, for which is currently estimated to be around 3\%~\cite{LHCHiggsCrossSectionWorkingGroup:2016ypw}, and those that affect the inclusive cross sections of \ggF \HH and single \PH production.

Sources that affect the cross sections include the PDFs, the strong coupling constant (\alpS), and variations of the renormalization ($\mu_\mathrm{R}$) and factorization ($\mu_\mathrm{F}$) scales, which account for missing higher-order terms in the perturbative calculations of the cross sections. Additionally, for the \ggF \HH production cross section, there is a source of uncertainty related to the scheme and scale choice of the top quark mass~\cite{Baglio:2020wgt}. These sources may also affect the kinematical variables of simulated events, thereby changing signal efficiencies. However, this effect has been found to be significantly smaller than the cross section uncertainty in recent \HGG~measurements~\cite{CMS:2021kom} and is therefore not considered.

The magnitudes of the uncertainties in the inclusive cross sections from these sources are taken from Refs.~\cite{Grazzini:2018bsd,Dawson:1998py,Borowka:2016ehy,Baglio:2018lrj,deFlorian:2013jea,Shao:2013bz,deFlorian:2015moa} for \ggF \HH production and from Refs.~\cite{pdf4lhc,LHCHiggsCrossSectionWorkingGroup:2013rie,LHCHiggsCrossSectionWorkingGroup:2016ypw} for single \PH boson production. The total effect of uncertainties on \HH production is found to be $^{+6.7\%}_{-23.2\%}$ for $\kl=1$, while the effect on resonant background production is between 2\% and 11\%, depending on the production mode.

\subsection{Experimental uncertainties}
As described previously, the sources of experimental uncertainty are split into two categories: those modifying the shape of the \mgg distribution (incorporated as nuisance parameters) and those affecting only the normalization (incorporated as log-normal uncertainties). This includes effects that lead to a migration of a process across analysis categories. Sources of experimental uncertainty that modify the shape of the \mgg distribution are related to the corrections applied to the photon energy scale and resolution in simulation. These corrections, and associated uncertainties, are computed using $\PZ\to\Pe\Pe$ events~\cite{CMS:2020uim}. The most dominant sources of experimental uncertainty, which affect only the normalization of signal and resonant background processes are:
\begin{itemize}
    \item \textit{Integrated luminosity}: The integrated luminosities for the 2016, 2017, and 2018 data-taking years have individual uncertainties of 1.2--2.5\%~\cite{CMS:2021xjt,CMS-PAS-LUM-17-004,CMS-PAS-LUM-18-002}, while the overall uncertainty in the measurement for the 2016--2018 period is 1.6\%. The correlations from different common sources of luminosity measurements are taken into account.
    \item \textit{Trigger efficiency}: This efficiency is calculated with the ``tag-and-probe'' method using $\PZ \to \Pe\Pe$ events in data~\cite{CMS:2011aa}. The size of uncertainty in the efficiency is between 0.1 and 3.5\% depending on the signal process and category. An additional uncertainty is also added to include the trigger inefficiency arising from a gradual shift in the timing of the inputs of the ECAL L1 trigger for $\abs{\eta} > 2.0$ in 2016--2017 data-taking years~\cite{CMS:2020cmk}. The size of this uncertainty is between 2 and 8\%.
    \item \textit{Diphoton preselection}: The efficiency after photon preselections is measured with simulated events and data, also using the ``tag-and-probe'' method with $\PZ \to \Pe\Pe$ events~\cite{CMS:2011aa}. The ratio of the efficiencies from data and MC simulations is used to derive the correction factor and corresponding uncertainties. The size of the uncertainty in the efficiency is between 4 and 5\%.
    \item \textit{b jet tagging}: The \textsc{DeepJet} discriminant distributions in MC simulations are reshaped by data-to-simulation scale factors~\cite{CMS:2017wtu}. The impact of different systematic uncertainty sources on the reshaping is studied. The most important effects lead to efficiency uncertainties up to 10\%.
    \item \textit{\tauh identification}: Selections are made using the \textsc{DeepTau} algorithm to distinguish genuine \tauh from jets, muons, and electrons. Measurements of the selection efficiency are made in data (as discussed in Section~\ref{sec:detector}) and are used to correct simulation. Uncertainties associated with those corrections are propagated into an uncertainty in the signal efficiency, which is found to be up to 7\%.
    \item \textit{Signal model interpolation}: There is an uncertainty related to the interpolation required to calculate signal efficiencies for the intermediate mass points. For each spline used, an alternative spline is considered where the nominal mass point closest to the intermediate mass point is removed. The relative difference between the prediction of the nominal and alternative spline is taken as an uncertainty in the efficiency of the signal within that category. 
\end{itemize}
Further sources of systematic uncertainty are included but are negligible compared to the dominant sources described above. In the final fits to data, the nuisance parameters are found to be consistent with their prefit values, and the combined impact of all systematic uncertainties is found to be less than 1\% on any of the upper limits reported in Section~\ref{sec:results}. Overall, the sensitivity for this search is limited mostly by the statistical uncertainty, given the low data yields in the sensitive analysis categories.

\section{Results\label{sec:results}}
\begin{figure}
    \centering
    \includegraphics[width=.4\linewidth]{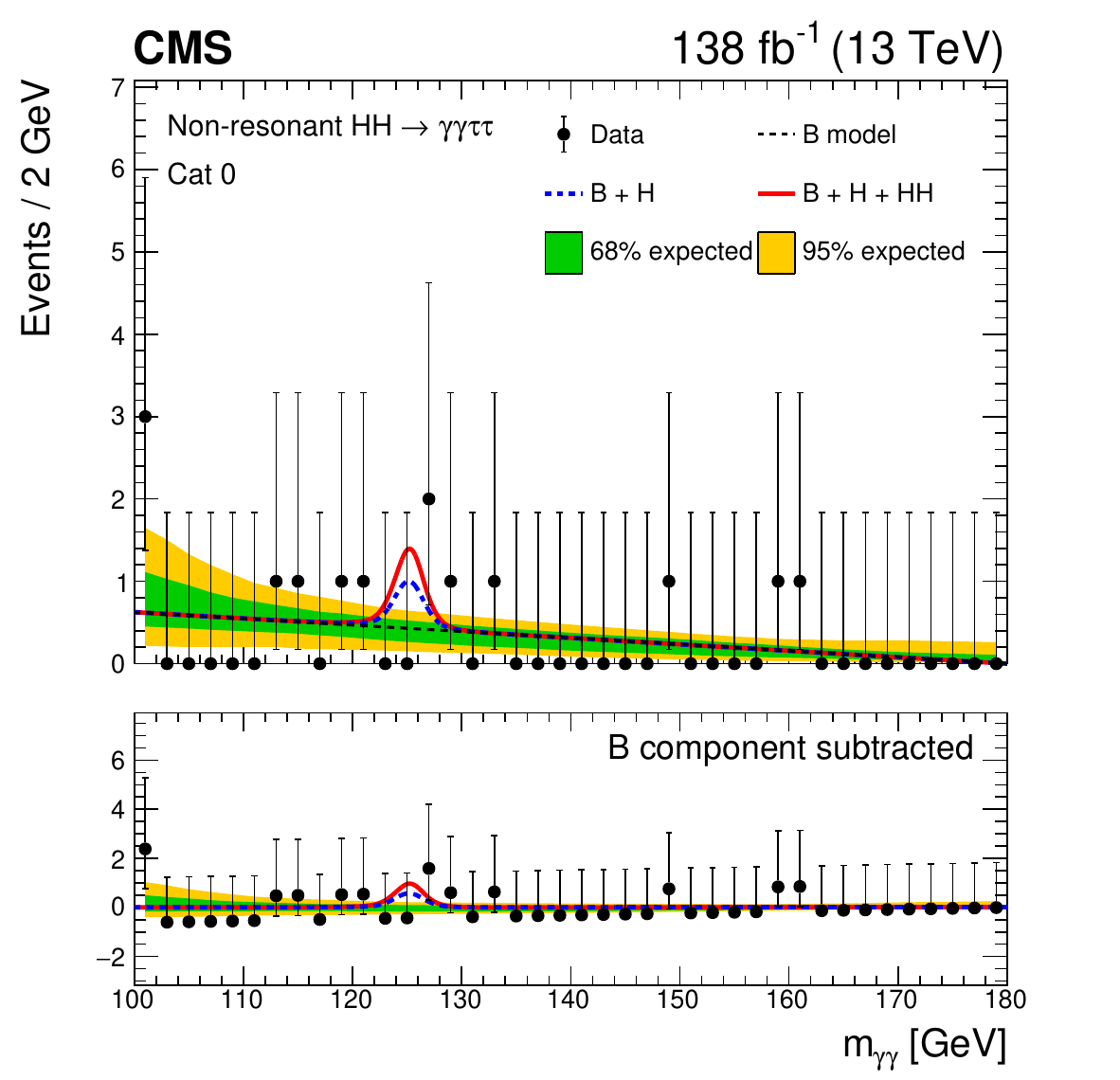}
    \includegraphics[width=.4\linewidth]{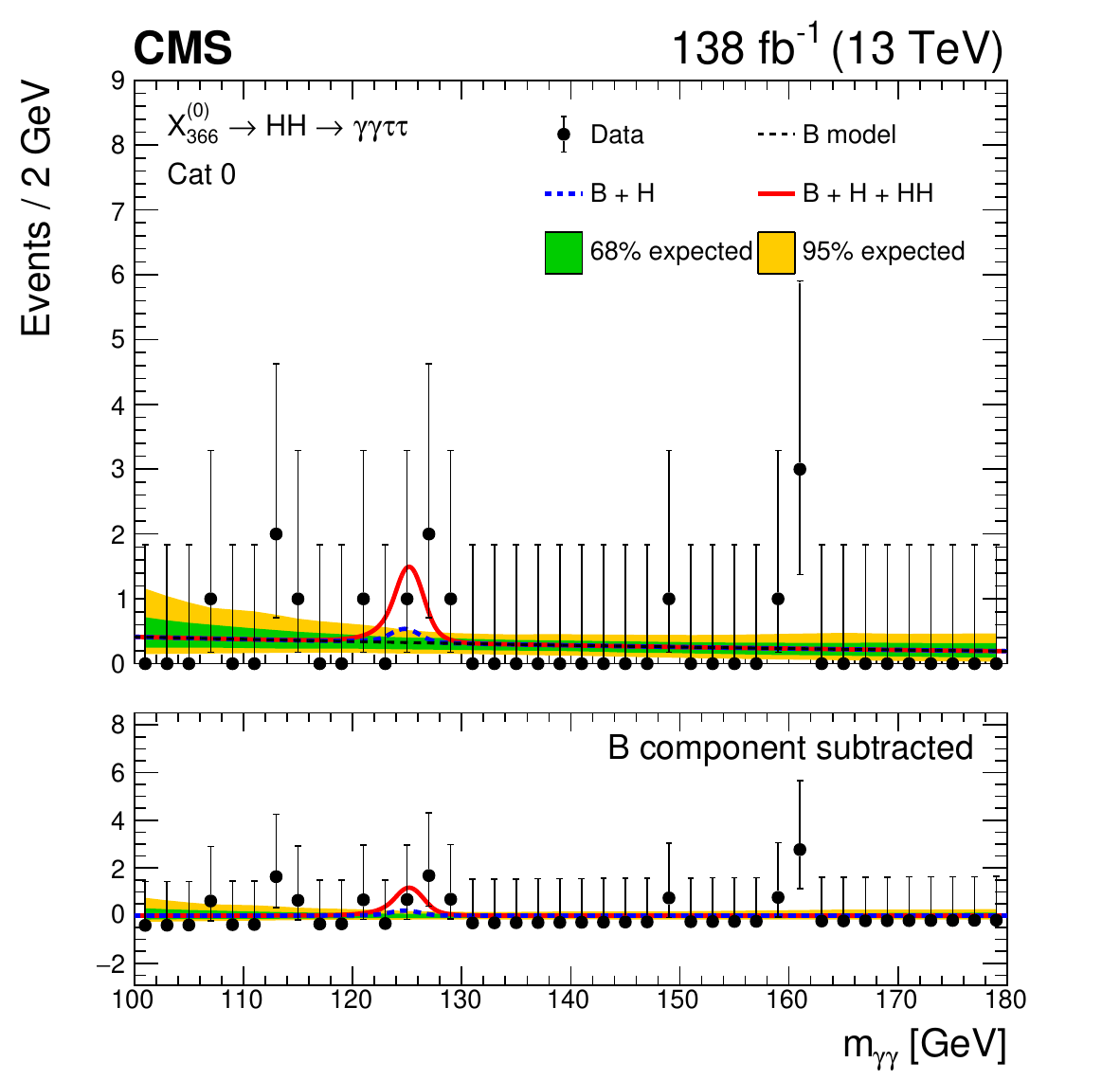}
    \includegraphics[width=.4\linewidth]{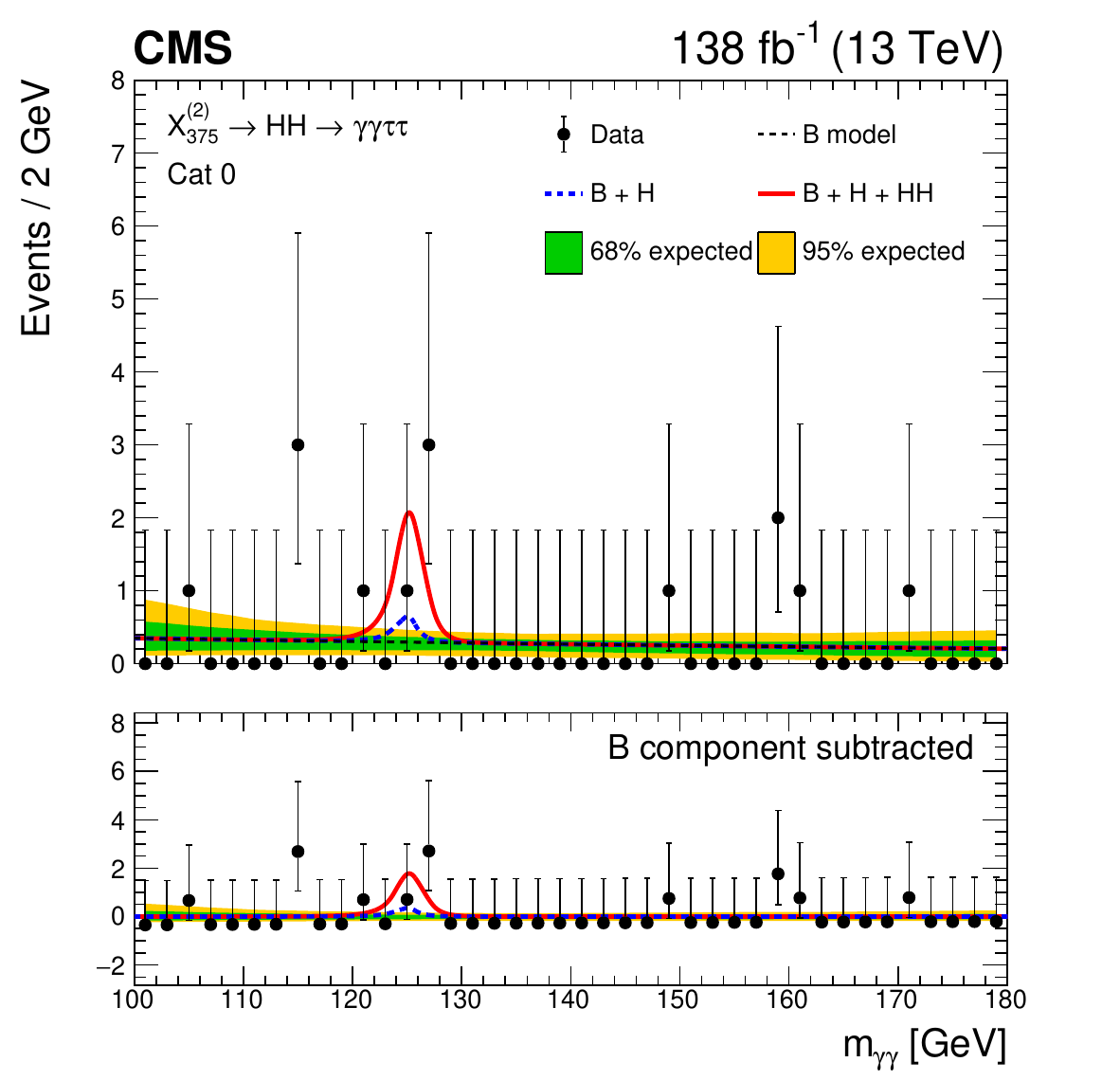}
    \includegraphics[width=.4\linewidth]{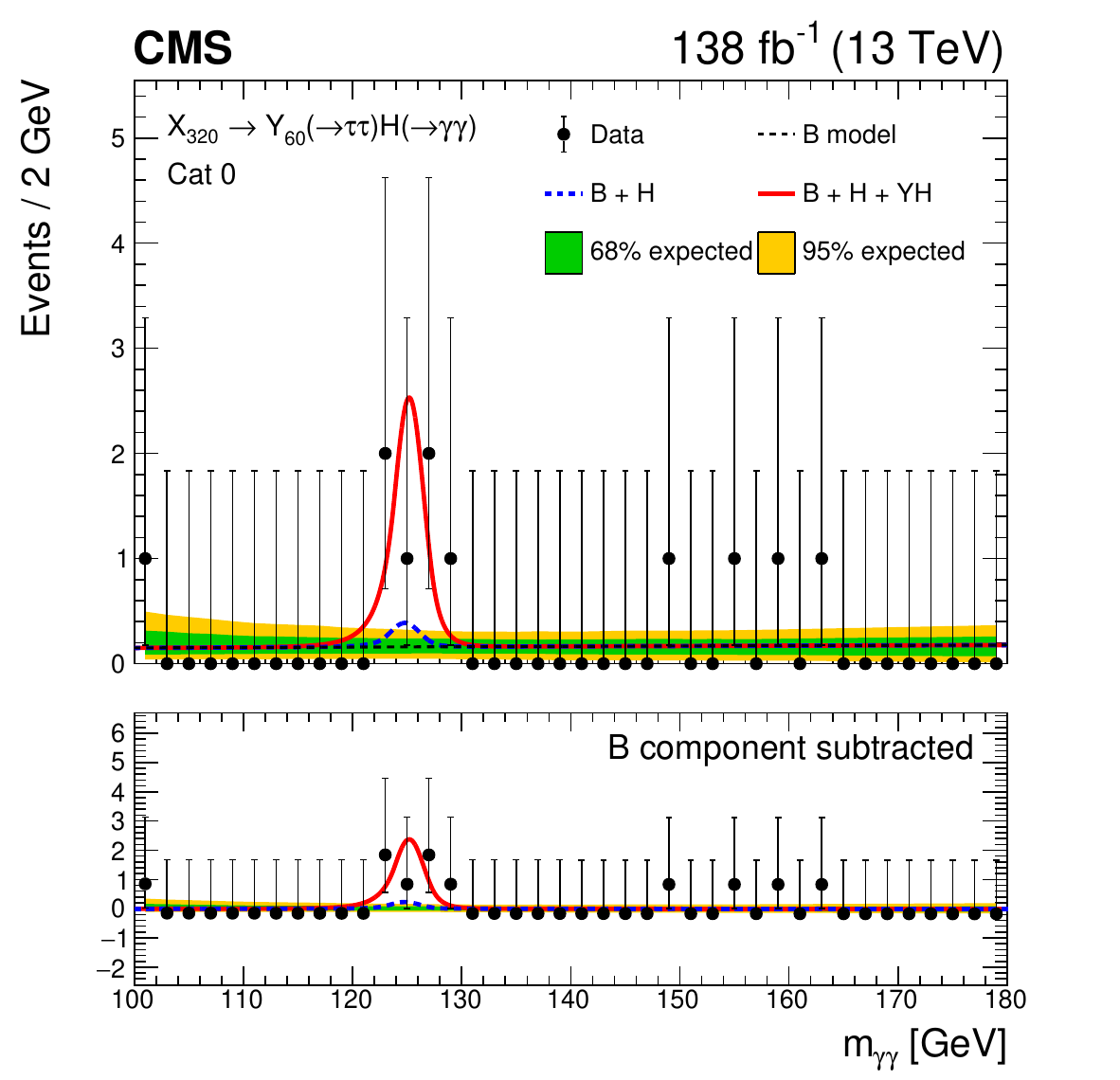}
    \includegraphics[width=.4\linewidth]{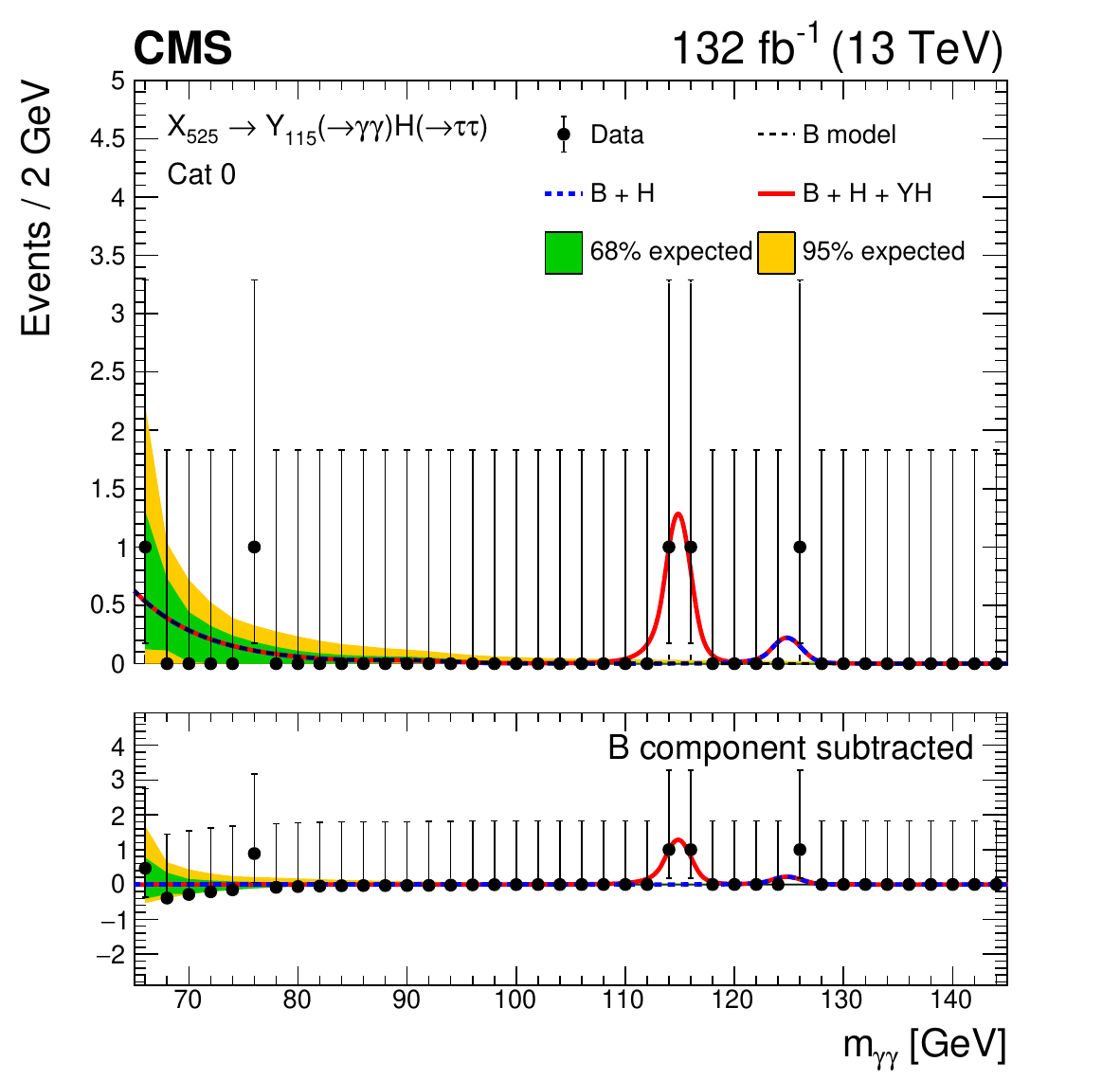}
    \includegraphics[width=.4\linewidth]{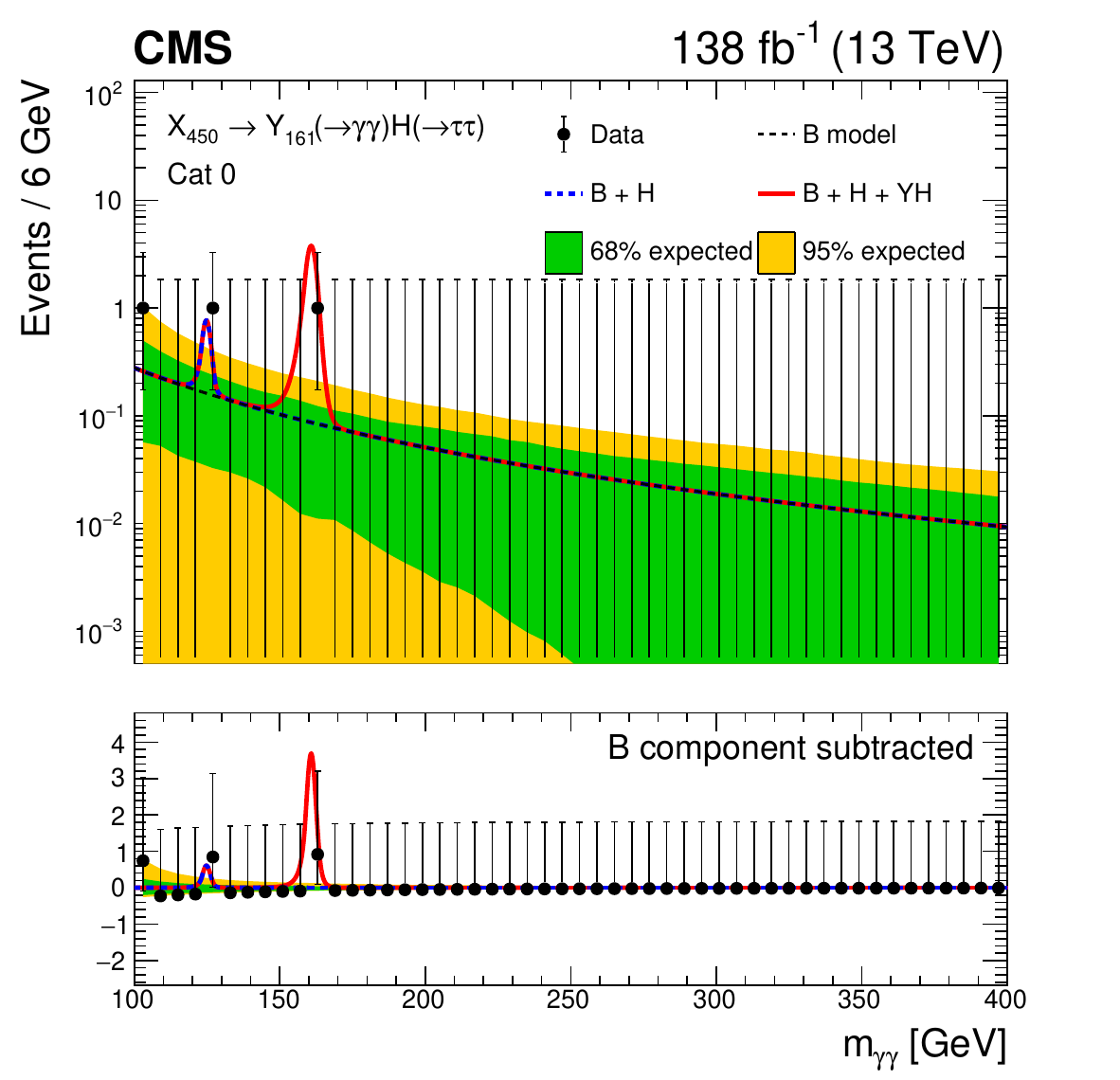}
    \caption{Data points (black) and signal-plus-background models for the most sensitive analysis category in each search are shown: nonresonant (upper left), \XZeroHH (upper right), \XTwoHH (middle left), \XYttHgg (middle right), low-mass \XYggHtt (lower left) and the high-mass \XYggHtt (lower right). The analysis categories for the resonant searches correspond to the mass hypotheses where the largest excesses with respect to the background-only hypothesis are observed. The one (green) standard deviation and two (yellow) standard deviation bands show the uncertainties in the background component of the fit. The solid red line shows the sum of the fitted signal and background components, the dashed blue line shows the continuum background and the background from single \PH production together, and the dashed black line shows only the continuum background component. The lower panel in each plot shows the residual signal yield after subtraction of the background. \label{fig:sbmodels}}
\end{figure}

For each search, a simultaneous maximum likelihood fit is performed to the \mgg distributions in data in the corresponding analysis categories. In the resonant searches, a separate fit is performed for each considered mass hypothesis. The best fit signal-plus-background models are shown for the best analysis categories in each search in Fig.~\ref{fig:sbmodels}. The figures for the resonant searches show the analysis categories for the \mx and (\mx, \my) points where the largest excesses with respect to the background-only hypothesis are observed.

\subsection{Results from the nonresonant search}
In the nonresonant \HH search, an upper limit at 95\% \CL is set on the production cross section of a pair of Higgs bosons, $\sigma({\Pp\Pp} \to \HH)$, using the \CLs modified frequentist criterion~\cite{CLS2,CLS1}, taking the LHC profile likelihood ratio as the test statistic~\cite{LHC-HCG}, in the asymptotic approximation~\cite{CLSA}. The observed (expected) 95\% \CL upper limit amounts to 930 (740)\unit{fb}, which corresponds to 33 (26) times the SM prediction.

\begin{figure}
    \centering
    \includegraphics[width=.7\linewidth]{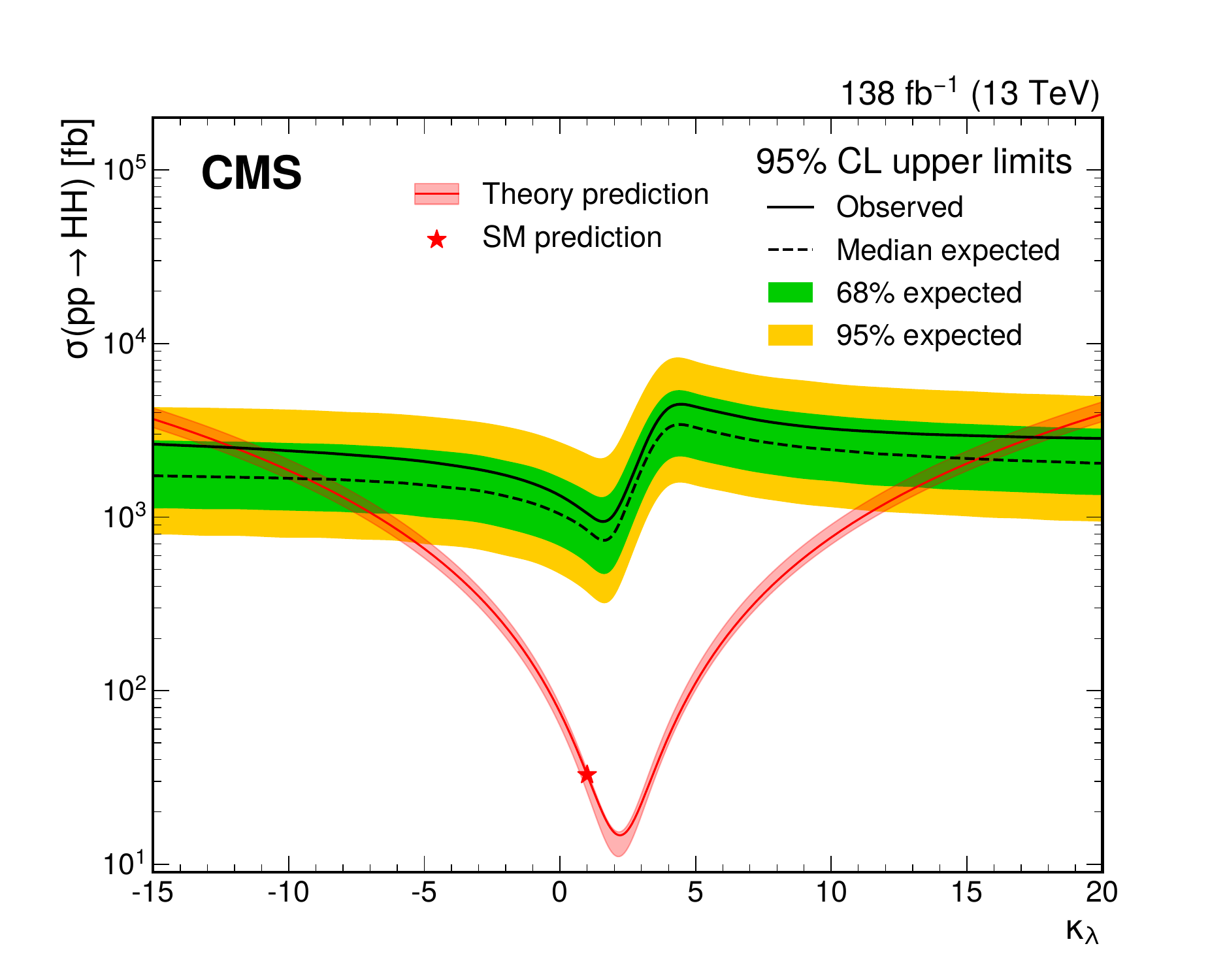}
    \caption{Expected and observed upper limits on the nonresonant \HH production cross section at 95\% \CL, obtained for different values of \kl. The green and yellow bands represent the one and two standard deviations for the expected limit, respectively. The theoretical prediction with the uncertainty of the cross section as a function of \kl is shown by the red band. \label{fig:limit_vs_klambda}}
\end{figure}

\begin{figure}
    \centering
    \includegraphics[width=.7\linewidth]{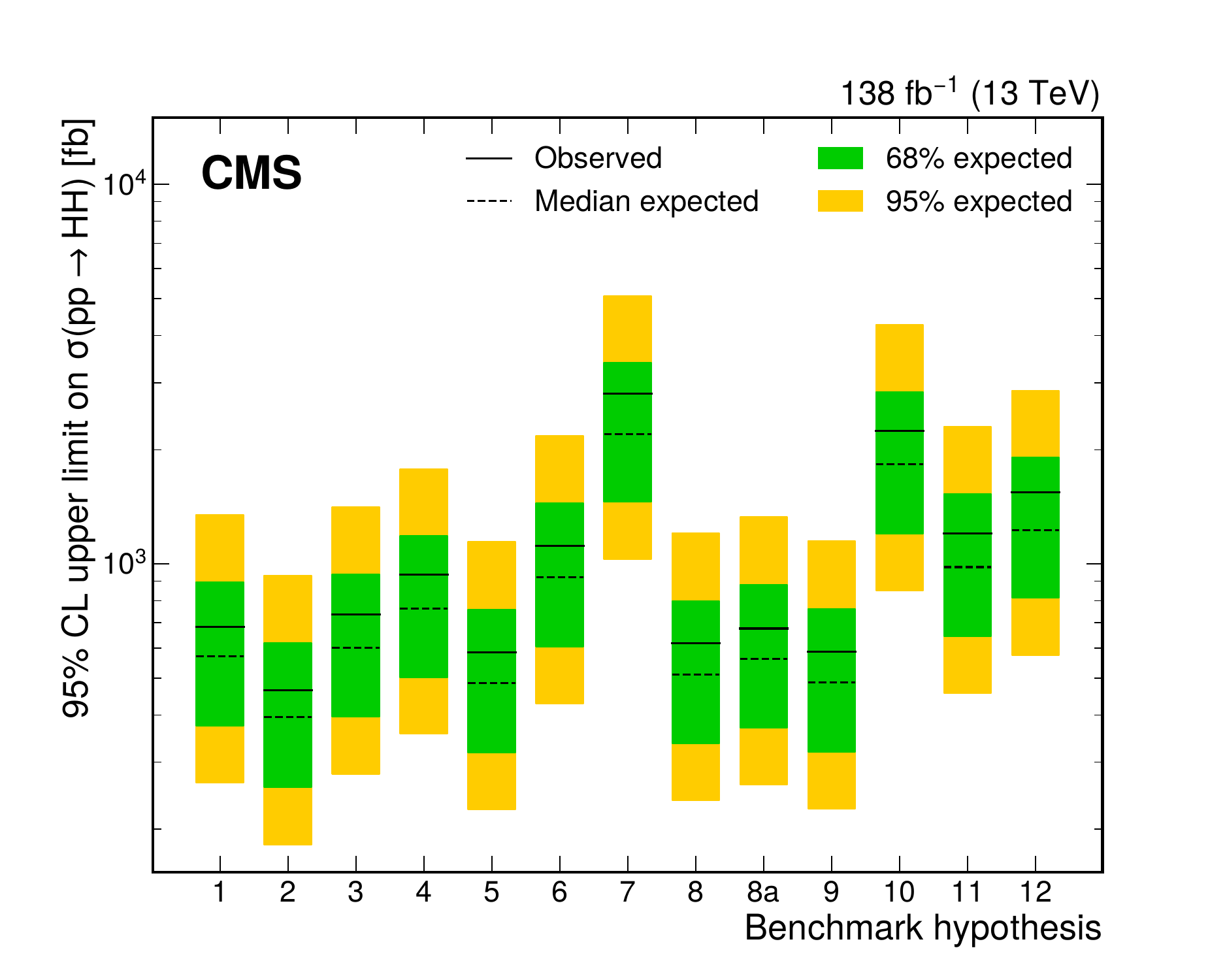}
    \caption{Expected and observed upper limits on the nonresonant \HH production cross section at 95\% \CL, for different thirteen BSM benchmark scenarios which consider different values of the couplings, $\kl$, $\kapt$, $\cgg$, $\cg$, and $\ctwo$ (defined in Table~\ref{tab:bench}). The green and yellow bands represent the one and two standard deviations for the expected limits, respectively. \label{fig:eft_benchmarks}}
\end{figure}

Limits are also derived for the nonresonant \HH search as a function of \kl, assuming that all other couplings of the Higgs boson have values as predicted by the SM. The result is shown in Fig.~\ref{fig:limit_vs_klambda}. The variation in the excluded cross section as a function of \kl arises due to the different kinematical properties of nonresonant \HH production for different values of \kl. At 95\% \CL, \HH production is observed (expected) to be excluded for values of \kl outside the range between $-12$ ($-9.4$) and 17 (15). The \kl-dependent NLO corrections in electroweak theory to single H production~\cite{Maltoni:2017ims} are neglected in this result. 
Upper limits at 95\% \CL are also set for the thirteen BSM benchmark scenarios listed in Table~\ref{tab:bench}. These limits are shown in Fig.~\ref{fig:eft_benchmarks}.

\subsection{Results from the resonant searches}
In the resonant searches, the limit-setting procedure is the same as in the nonresonant search except that the test statistic is not assumed to be distributed according to the asymptotic approximation. The results are extracted both in the asymptotic approximation and using pseudo-datasets. In case the results are not compatible, the ones using the pseudo-datasets are presented. This is most important in the high-mass \XYggHtt search where there are small numbers of events in the tail of the \mgg distribution. Upper limits are presented on production cross sections, factoring out the relevant branching fractions where known. The \XYttHgg search is an exception to this rule and the \HGG branching fraction is not factored out to allow for a more straightforward comparison to the \XYbbHgg search presented in Ref.~\cite{CMS:2023boe}.

\begin{figure}
    \centering
    \includegraphics[width=.7\textwidth]{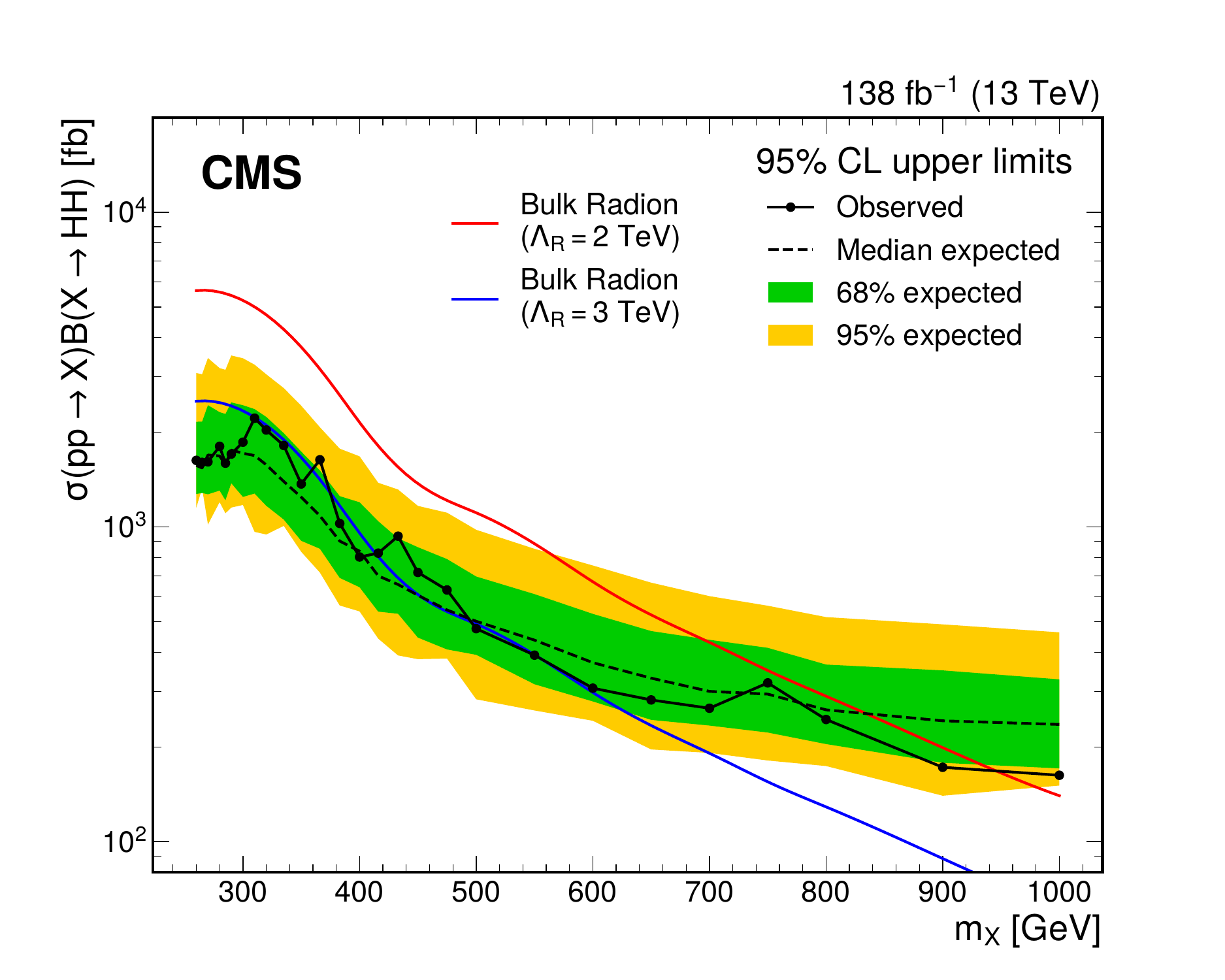}
    \includegraphics[width=.7\textwidth]{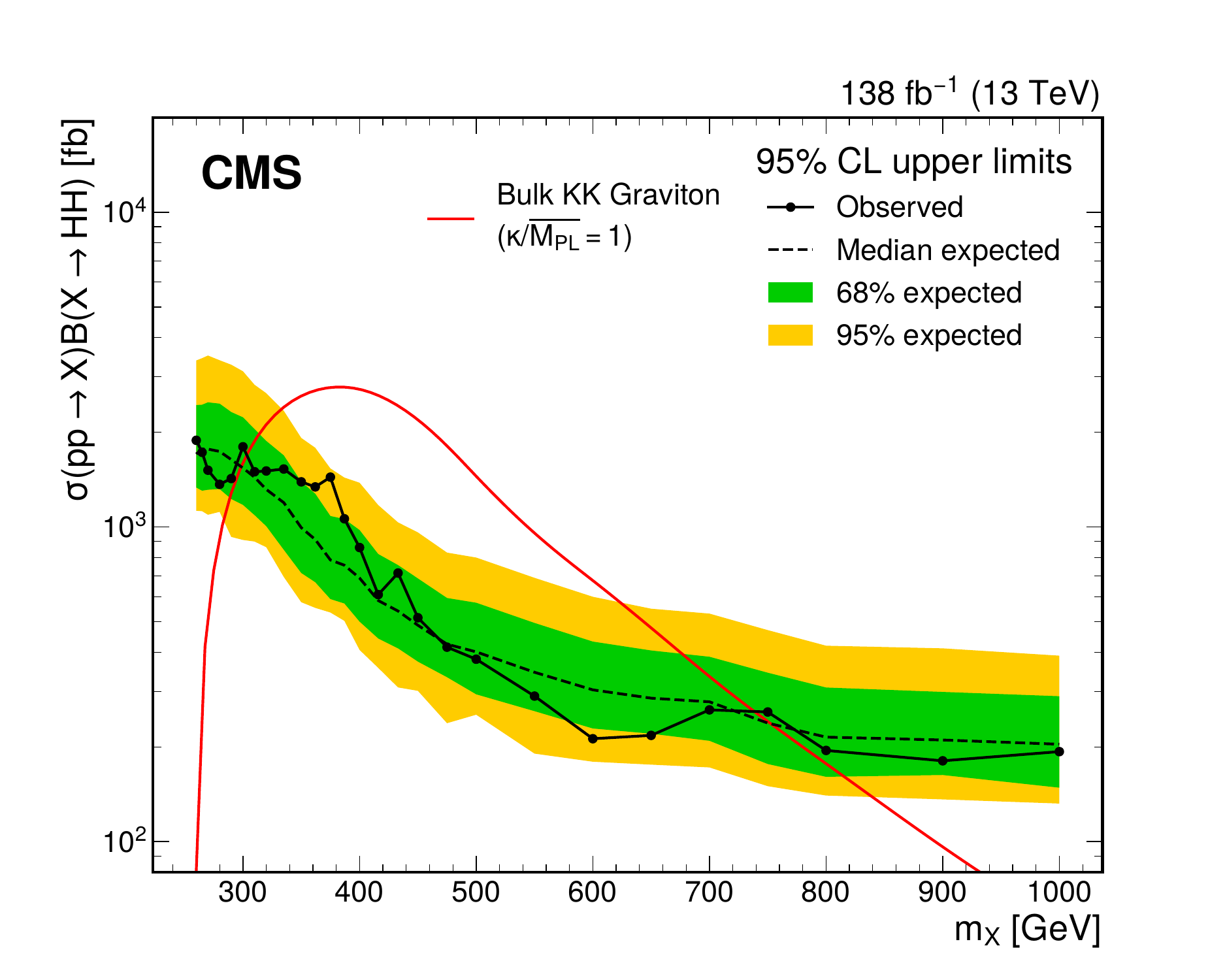}
    \caption{Expected and observed 95\% \CL upper limits on the resonant production cross section, $\sigma({\Pp\Pp} \to \PX)B(\PX \to \HH)$ for the spin-0 \XZeroHH search (upper plot) and spin-2 \XTwoHH search (lower plot). The dashed and solid black lines represent the expected and observed limits, respectively. The green and yellow bands represent the one and two standard deviations for the expected limit, respectively. The red and blue lines show the theoretical predictions with different energy scales and couplings~\cite{Carvalho:2014lsg}. \label{fig:limits_xhh}}
\end{figure}

Upper limits on the resonant production cross section, $\sigma({\Pp\Pp} \to \PX)B(\PX \to \HH)$, are shown in Fig.~\ref{fig:limits_xhh} as a function of the resonant mass \mx, for the \XZeroHH (upper plot) and \XTwoHH (lower plot) searches. The expected limits are shown to decrease as a function of \mx, which follows from the improved signal \vs background discrimination for high \mx values. The observed (expected) upper limits vary between 160--2200\unit{fb} (240--1800\unit{fb}) and 180--1900\unit{fb} (200--1800\unit{fb}) for the \XZeroHH and \XTwoHH searches, respectively. In the context of the RS bulk model, these results exclude \mx values between 260 and 360\GeV, between 380 and 410\GeV, and between 500 and 550\GeV for the spin-0 radion resonance with decay constant $\Lambda_\mathrm{R}=3$\TeV, and up to 950\GeV at $\Lambda_\mathrm{R}=2$\TeV. For the spin-2 KK graviton resonance with coupling factor $\kappa/\overline{M_\mathrm{pl}}=1$, masses between 310 and 750\GeV are excluded.

\begin{figure}
    \centering
    \includegraphics[width=1.0\textwidth]{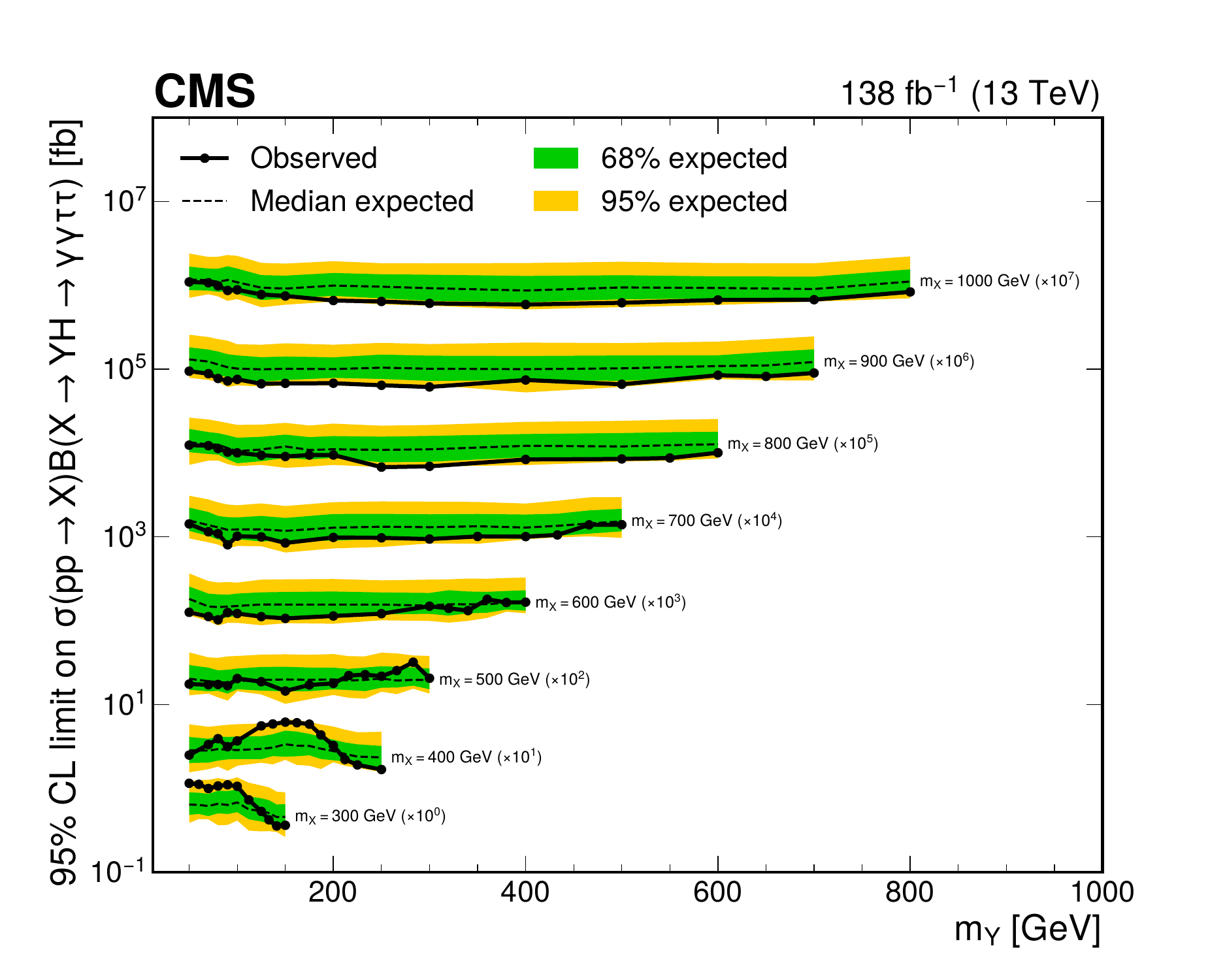}
    \caption{Expected and observed 95\% \CL upper limits on $\sigma({\Pp\Pp} \to \PX)B(\PX \to \YH \to \ggtt)$ for the \XYttHgg search as a function as a function of \my for different \mx hypotheses. The dashed and solid black lines represent the expected and observed limits, respectively. The green and yellow bands represent the one and two standard deviations for the expected limit, respectively. Limits are scaled by orders of 10, labeled in the plot. \label{fig:limits_xytthgg}}
\end{figure}

\begin{figure}
    \centering
    \includegraphics[width=1.0\textwidth]{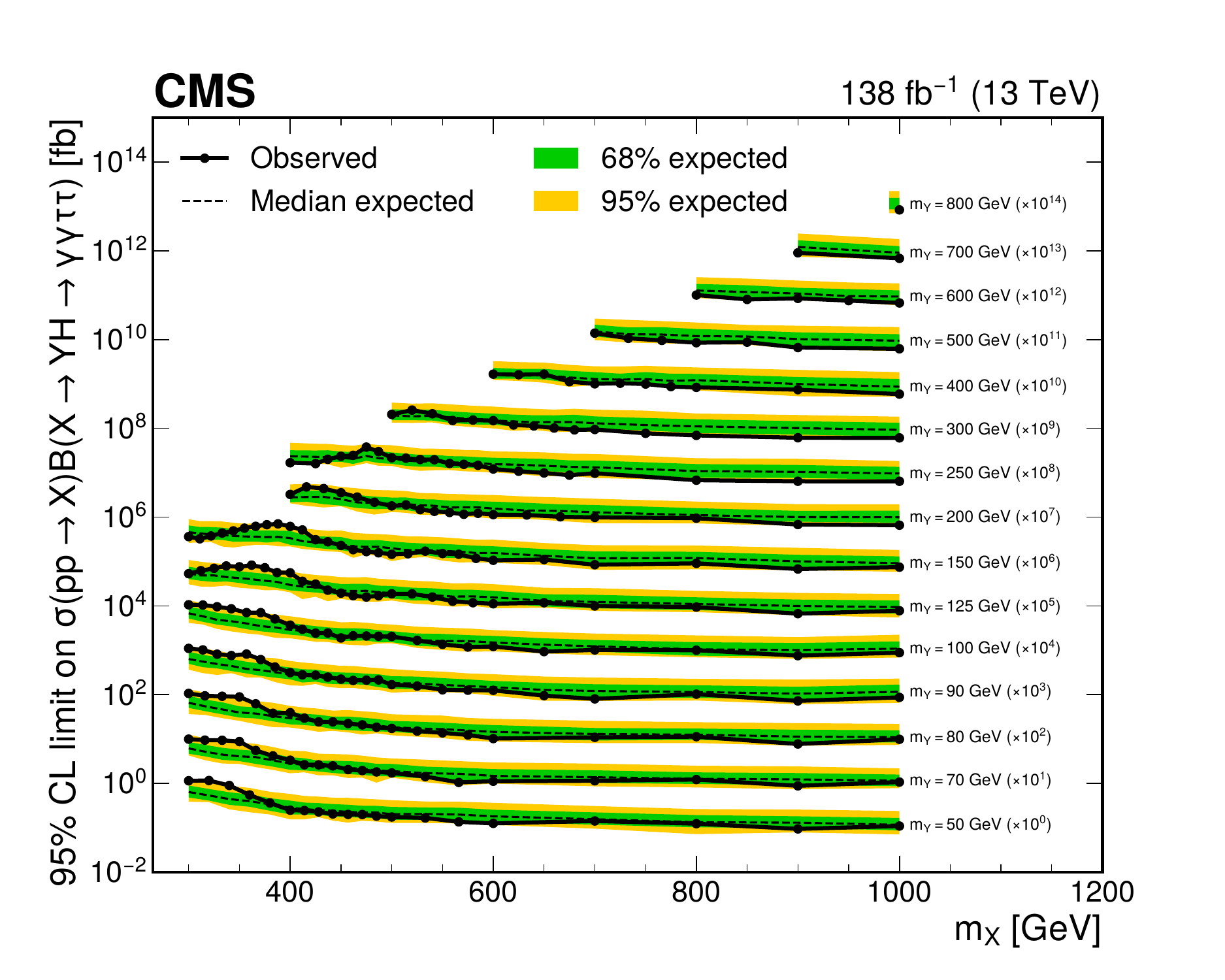}
    \caption{Expected and observed 95\% \CL upper limits on $\sigma({\Pp\Pp} \to \PX)B(\PX \to \YH \to \ggtt)$ for the \XYttHgg search as a function of \mx for different \my hypotheses. The dashed and solid black lines represent the expected and observed limits, respectively. The green and yellow bands represent the one and two standard deviations for the expected limit, respectively. Limits are scaled by orders of 10, labeled in the plot. \label{fig:limits_xytthgg2}}
\end{figure}

\begin{figure}
    \centering
    \includegraphics[width=1.0\textwidth]{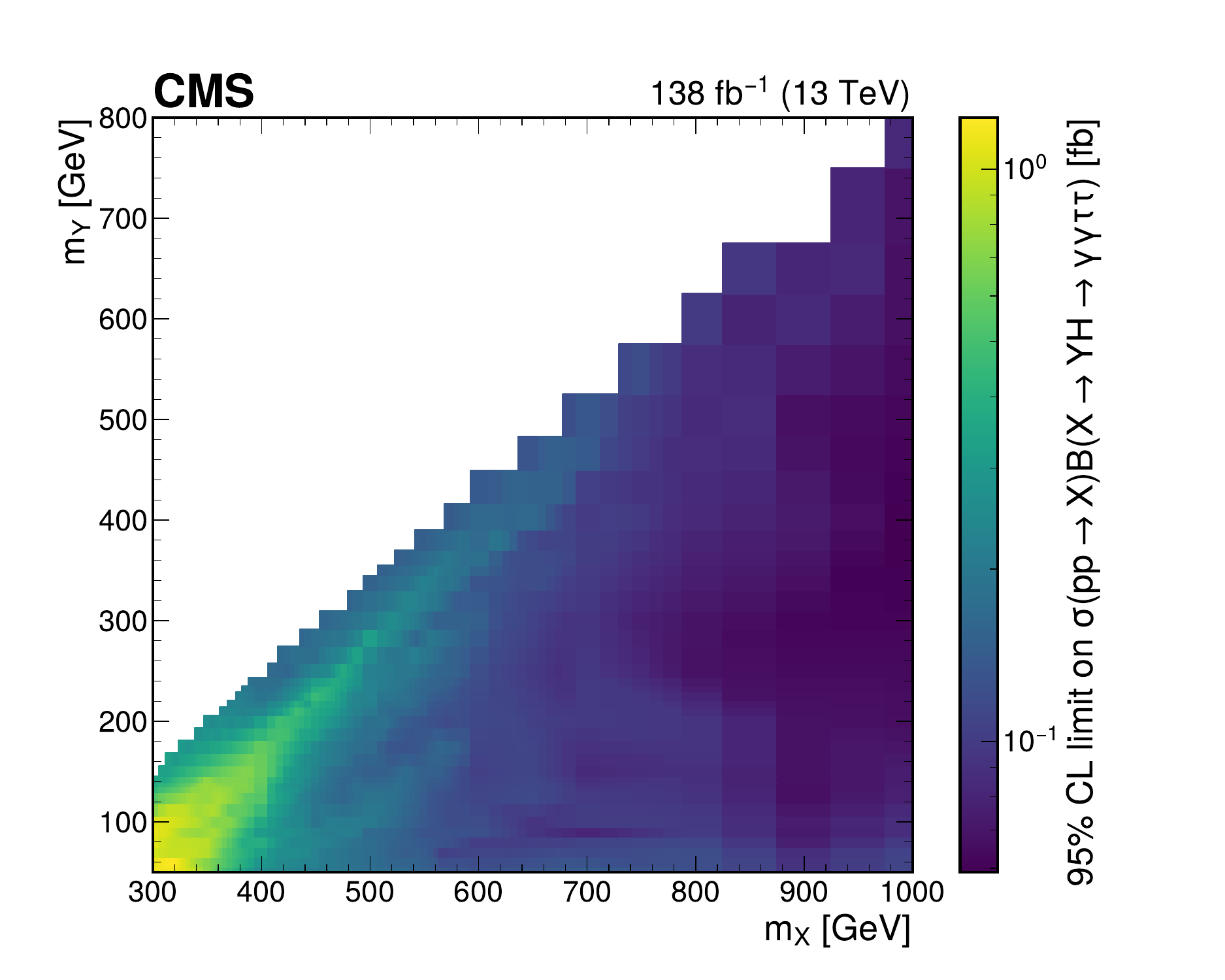}
    \caption{Observed upper limits in the 2D (\mx, \my) plane for the \XYttHgg search. The values of the limits are shown by the color scale. \label{fig:limits_xytthgg3}}
\end{figure}

\begin{figure}
    \centering
    \includegraphics[width=1.0\textwidth]{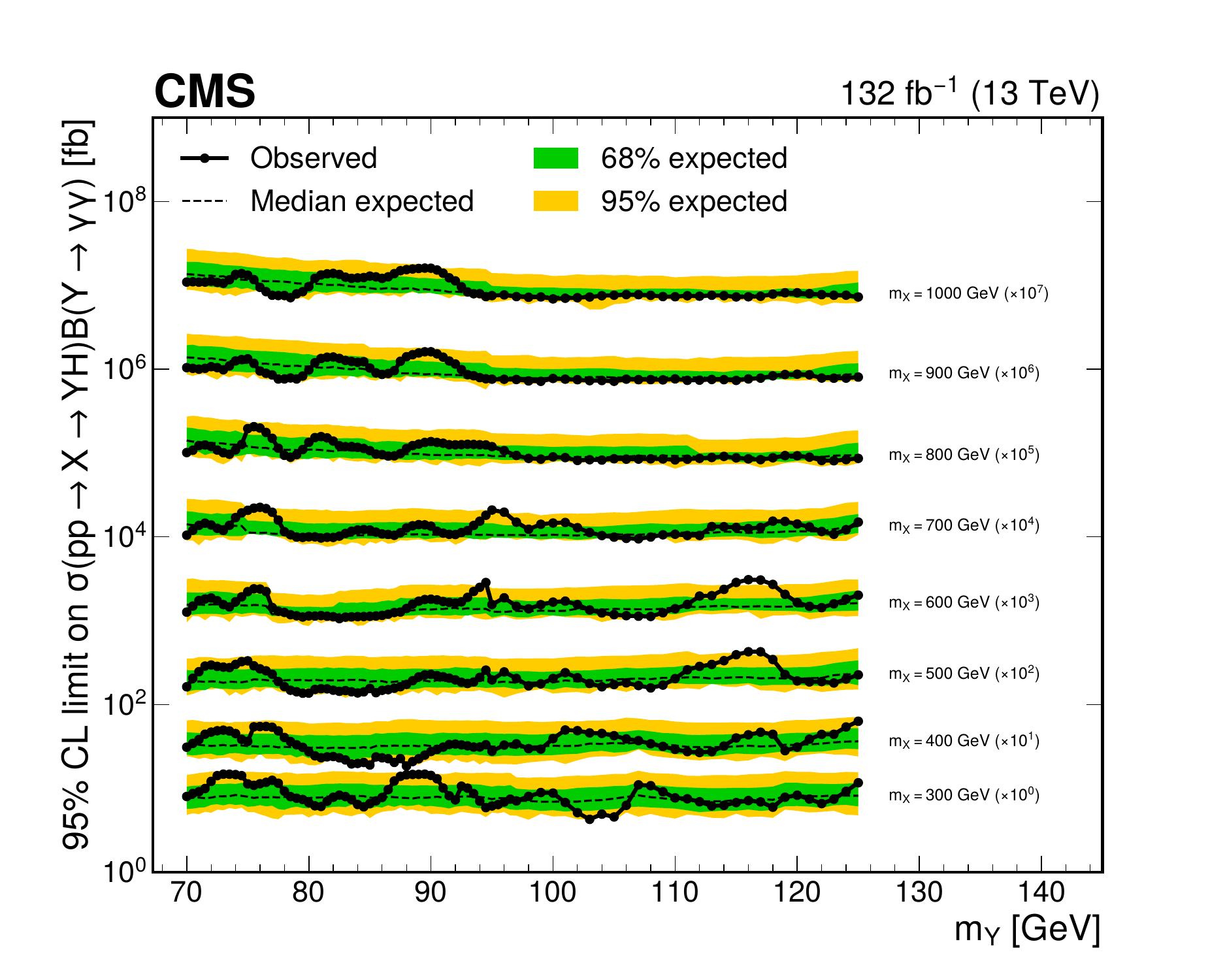}
    \caption{Expected and observed 95\% \CL upper limits on $\sigma({\Pp\Pp} \to \PX \to \YH)B(\PY \to \PGg\PGg)$ for the low-mass \XYggHtt search as a function of \my for different \mx hypotheses. The dashed and solid black lines represent the expected and observed limits, respectively. The green and yellow bands represent the one and two standard deviations for the expected limit, respectively. Limits are scaled by orders of 10, labeled in the plot. \label{fig:limits_xygghtt_lowmass}}
\end{figure}

\begin{figure}
    \centering
    \includegraphics[width=1.0\textwidth]{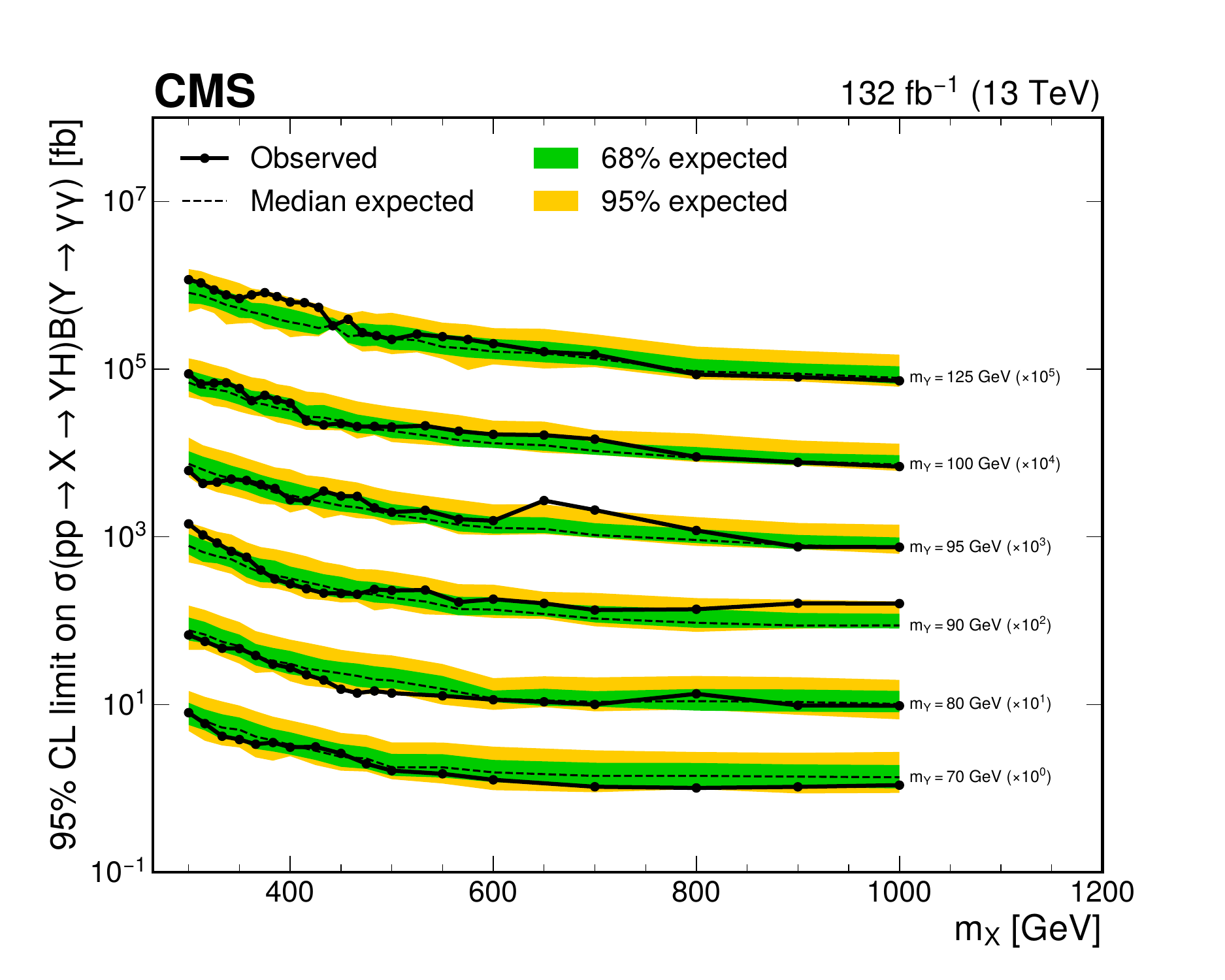}
    \caption{Expected and observed 95\% \CL upper limits on $\sigma({\Pp\Pp} \to \PX \to \YH)B(\PY \to \PGg\PGg)$ for the low-mass \XYggHtt search as a function of \mx for different \my hypotheses. The dashed and solid black lines represent the expected and observed limits, respectively. The green and yellow bands represent the one and two standard deviations for the expected limit, respectively. Limits are scaled by orders of 10, labeled in the plot. \label{fig:limits_xygghtt_lowmass2}}
\end{figure}

\begin{figure}
    \centering
    \includegraphics[width=1.0\textwidth]{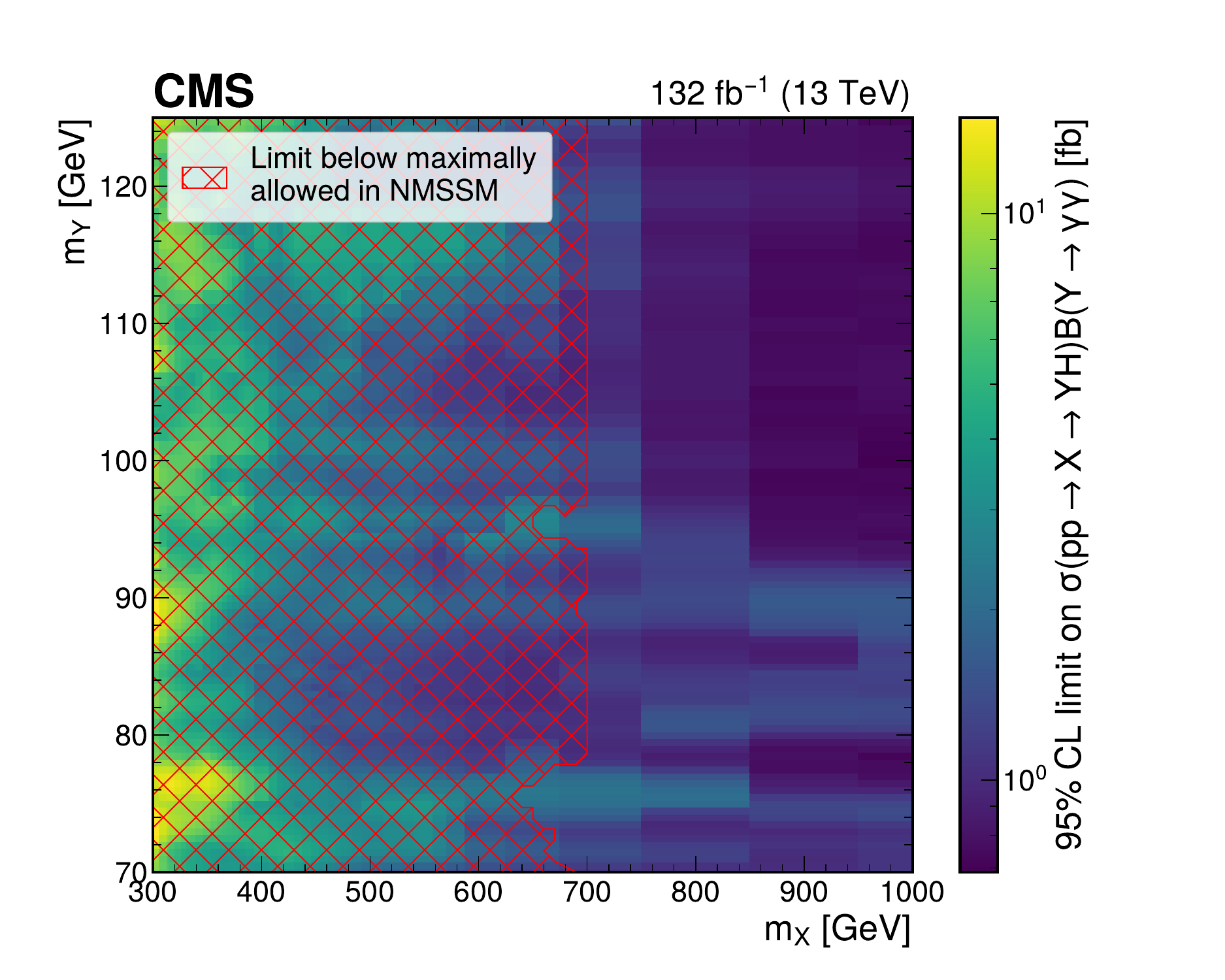}
    \caption{Observed upper limits in the 2D (\mx, \my) plane for the low-mass \XYggHtt search. The values of the limits are shown by the color scale. The red hatched region indicates masses for which the observed limits exclude largest possible cross-section in the NMSSM from previous constraints, taken from Ref.~\cite{Ellwanger:2022jtd}.\label{fig:limits_xygghtt_lowmass3}}
\end{figure}

\begin{figure}
    \centering
    \includegraphics[width=1.0\textwidth]{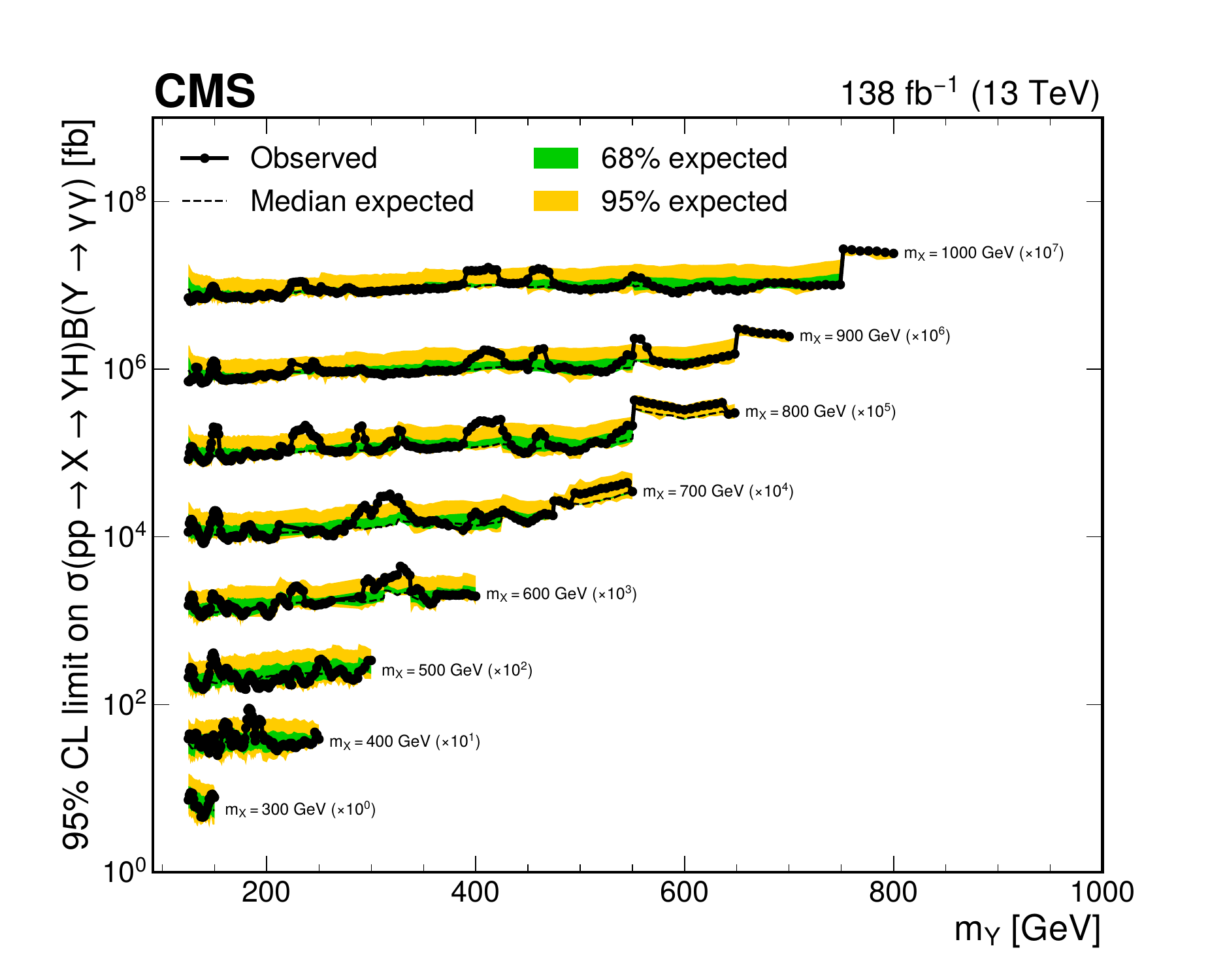}
    \caption{Expected and observed 95\% \CL upper limits on $\sigma({\Pp\Pp} \to \PX \to \YH)B(\PY \to \PGg\PGg)$ for the high-mass \XYggHtt search as a function of \my for different \mx hypotheses. The dashed and solid black lines represent the expected and observed limits, respectively. The green and yellow bands represent the one and two standard deviations for the expected limit, respectively. Limits are scaled by orders of 100, labeled in the plot. \label{fig:limits_xygghtt_highmass}}
\end{figure}

\begin{figure}
    \centering
    \includegraphics[width=1.0\textwidth]{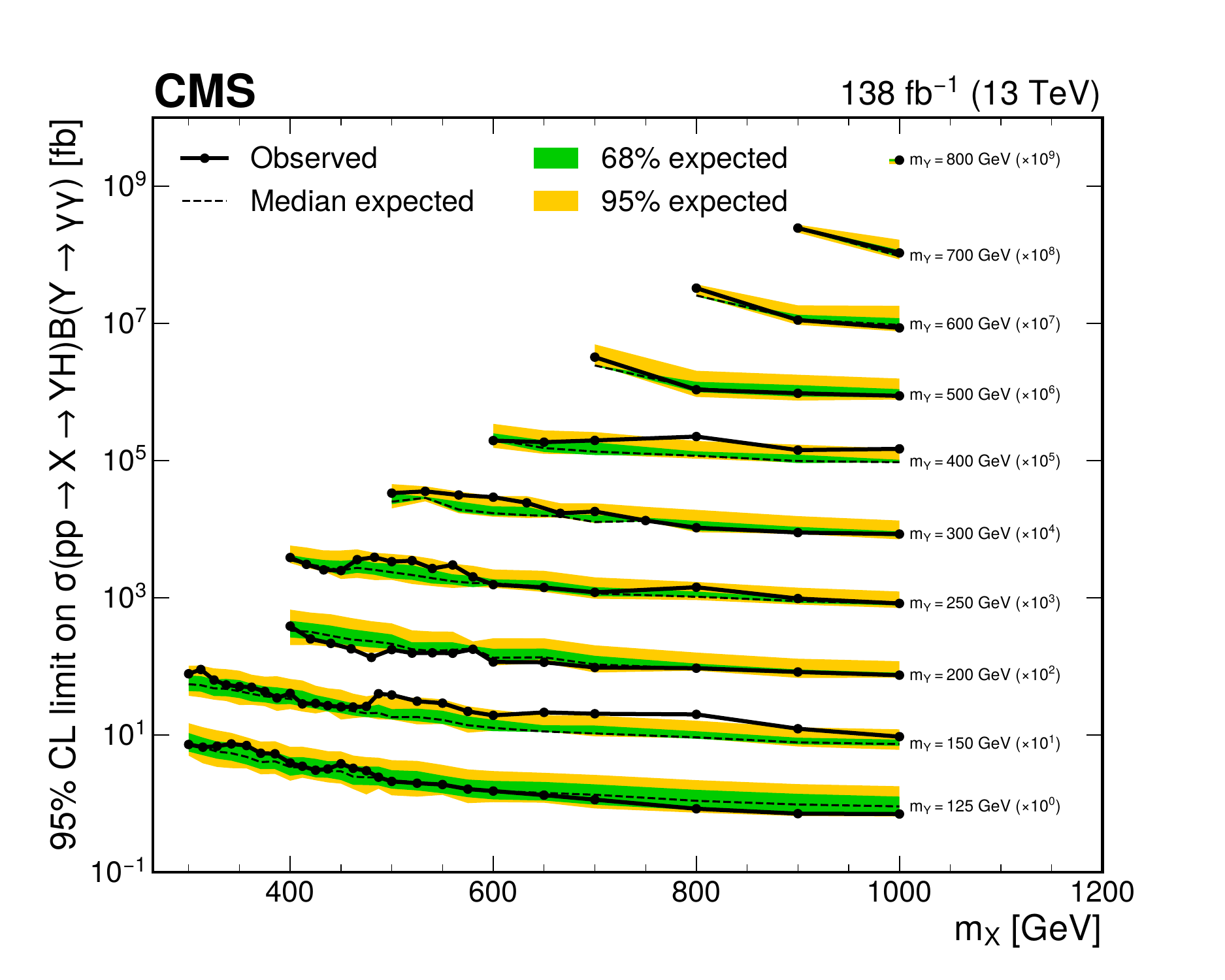}
    \caption{Expected and observed 95\% \CL upper limits on $\sigma({\Pp\Pp} \to \PX \to \YH)B(\PY \to \PGg\PGg)$ for the high-mass \XYggHtt search as a function of \mx for different \my hypotheses. The dashed and solid black lines represent the expected and observed limits, respectively. The green and yellow bands represent the one and two standard deviations for the expected limit, respectively. Limits are scaled by orders of 10, labeled in the plot.  \label{fig:limits_xygghtt_highmass2}}
\end{figure}

\begin{figure}
    \centering
    \includegraphics[width=1.0\textwidth]{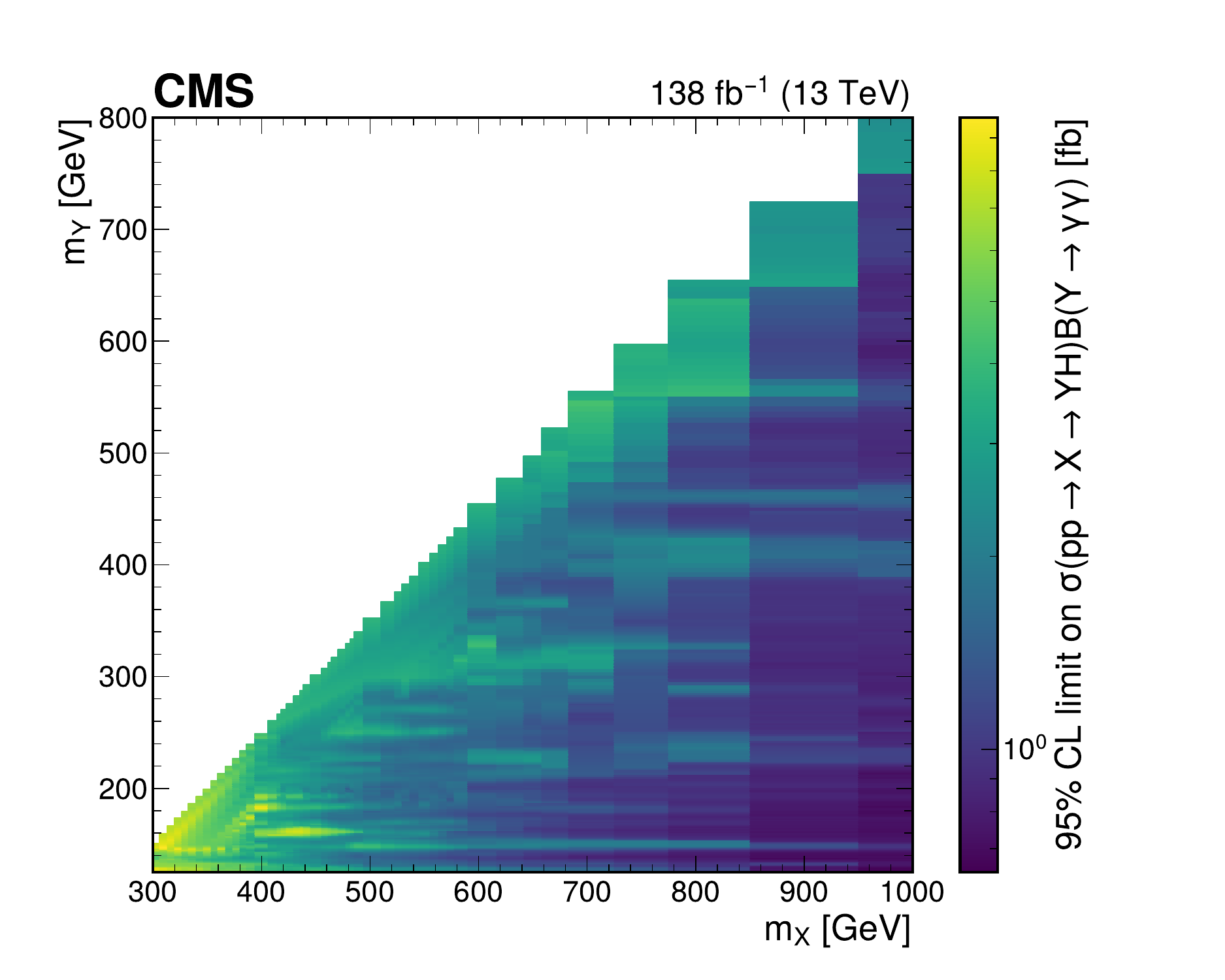}
    \caption{Observed upper limits in the 2D (\mx, \my) plane for the high-mass \XYggHtt search. The values of the limits are shown by the color scale. \label{fig:limits_xygghtt_highmass3}}
\end{figure}

The upper limits on the product of the production cross section and the decay branching fraction $\sigma({\Pp\Pp} \to \PX)B(\PX \to \YH \to \ggtt)$ for the \XYttHgg search are shown in Figs.~\ref{fig:limits_xytthgg},~\ref{fig:limits_xytthgg2}, and \ref{fig:limits_xytthgg3}. The limits are shown as a function of \my for different \mx hypotheses (Fig.~\ref{fig:limits_xytthgg}) or as a function of \mx for different \my hypotheses (Fig.~\ref{fig:limits_xytthgg2}). The observed upper limits in the 2D (\mx, \my) plane are also shown in Fig.~\ref{fig:limits_xytthgg3}. Analogously, the upper limits on the product of the production cross section and the decay branching fraction $\sigma({\Pp\Pp} \to \PX \to \YH)B(\PY \to \PGg\PGg)$ for the \XYggHtt searches are shown in Figs.~\ref{fig:limits_xygghtt_lowmass},~\ref{fig:limits_xygghtt_lowmass2}, and \ref{fig:limits_xygghtt_lowmass3} for the low-mass search, which considers \my in the range from 70 to 125\GeV, and in Figs.~\ref{fig:limits_xygghtt_highmass},~\ref{fig:limits_xygghtt_highmass2}, and \ref{fig:limits_xygghtt_highmass3} for the high-mass search, which considers \my in the range from 125 to 800\GeV. The limits are shown as a function of \my for different \mx hypotheses (Figs.~\ref{fig:limits_xygghtt_lowmass} and \ref{fig:limits_xygghtt_highmass}), as a function of \mx for different \my hypotheses (Figs.~\ref{fig:limits_xygghtt_lowmass2} and \ref{fig:limits_xygghtt_highmass2}), and in the 2D (\mx, \my) plane (Figs.~\ref{fig:limits_xygghtt_lowmass3} and \ref{fig:limits_xygghtt_highmass3}).

For the \XYttHgg search, the observed (expected) upper limits vary between  0.059--1.2\unit{fb} (0.087--0.68\unit{fb}), depending on the \mx and \my values. In the low-mass \XYggHtt search, the observed (expected) upper limits vary between 0.69--15\unit{fb} (0.73--8.3\unit{fb}), depending on the \mx and \my values. These limits can be compared to a set of maximally allowed cross sections in the NMSSM taken from Ref.~\cite{Ellwanger:2022jtd}, which are calculated using the \textsc{NMSSMTools} v5.6.2 package~\cite{Ellwanger:2004xm,Ellwanger:2005dv,Belanger:2013oya}. Owing to the allowance of a high \Ygg branching fraction, a region in (\mx, \my) parameter space is found where the observed limits are below the maximally allowed limits, implying that these results can be used to provide tighter constraints on the NMSSM. This region is shown by the red dashed area in Fig.~\ref{fig:limits_xygghtt_lowmass3}. For the high-mass \XYggHtt search, the observed (expected) upper limits vary between 0.64--10\unit{fb} (0.70--7.6\unit{fb}), depending on the \mx and \my values.

In the \XHH searches, no excesses with local significances above 1.7 standard deviations are found. In the \XYttHgg search, the highest local significance seen is 2.6 standard deviations at (\mx, \my) = (320, 60)\GeV. In the low-mass and high-mass \XYggHtt searches, the highest local significances found are 3.2 standard deviations at (\mx, \my) = (525, 115)\GeV, and 3.2 standard deviations at (\mx, \my) = (450, 161)\GeV respectively. Due to the recent CMS excesses discussed in Section~\ref{sec:intro} which are consistent with resonances of masses 650\GeV and 90--100\GeV, excesses at similar masses are also worth reporting. No significant excess is seen in the \XYttHgg search at these masses, but in the \XYggHtt search, local significances of 2.2 and 2.3 standard deviations are seen at (\mx, \my) = (600,94.5)\GeV and (\mx, \my) = (650,95)\GeV respectively. The significance values have been determined for a finer granularity of mass points than that used for the presentation of upper limits. The largest excesses are therefore not necessarily clearly visible in the figures presented. 

Due to the large range in mass hypotheses considered and the narrow resolution of the signal peak, there is a large look-elsewhere effect. The global significance, determined as the probability to observe an excess at any of the mass points considered in the scans that is at least as large as the largest one observed in data, is computed by running multiple pseudo-experiments. The global significances for the \XYttHgg, low-mass \XYggHtt, and high-mass \XYggHtt searches are 2.2, 0.1, and 0.3 standard deviations, respectively. Therefore, as a standalone analysis, these searches do not show significant deviation from the SM. However, the excesses at $\my = 95\GeV$ are interesting since they are consistent with other excesses reported by CMS, and they provide further motivation to perform measurements at these mass points in the future.

\section{Summary\label{sec:summary}}
A search for the production of two scalar bosons in the \ggtt final state is presented. The search uses data from proton-proton collisions collected by the CMS experiment at the LHC in 2016--2018 at a center-of-mass energy of 13\TeV, corresponding to 138 or 132\fbinv of integrated luminosity, depending on the trigger path. In total, five searches are performed. One search targets the nonresonant production of a Higgs boson pair, \HH, via gluon-gluon fusion, where no significant deviation from the background-only hypothesis is observed. Upper limits at 95\% confidence level (CL) on the \HH production cross section are extracted for production in the standard model (SM) and in several beyond-the-SM scenarios. The observed (expected) upper limit for the SM production is found to be 930 (740)\unit{fb}, corresponding to 33 (26) times the SM prediction. The limit is also derived as a function of the Higgs boson self-coupling modifier, \kl, assuming all other Higgs boson couplings are as predicted in the SM, and \HH production is observed (expected) to be excluded at 95\% \CL outside the range between $-12$ ($-9.4$) and 17 (15). In addition, the limit is extracted for numerous beyond-the-SM benchmark scenarios and the results are consistent with the SM predictions.

This analysis also targets the resonant production of two scalar bosons. Two searches are constructed to search for a resonance \PX decaying to a SM Higgs boson pair, \XHH, for both the spin-0 resonance and spin-2 resonance scenarios. Furthermore, the analysis targets the \XYH process, where \PY is an additional, lighter (than X) scalar particle. Three searches are constructed, namely the \XYttHgg, low-mass \XYggHtt and high-mass \XYggHtt, which target the different decay chains in different mass regimes. The global significances of each search indicate that no significant deviation from the background-only hypothesis is observed, but some local excesses in the \XYggHtt search that are consistent with other CMS results warrant further measurements. 

In the \XHH search, upper limits at the 95\% \CL on the \XHH production cross section are observed (expected) to be within 160 to 2200 (200 to 1800)\unit{fb}, depending on the mass of \PX. In the \XYttHgg search, upper limits at the 95\% \CL on the product of the production cross section and the decay branching fractions are observed (expected) to vary between 0.059--1.2\unit{fb} (0.087--0.68\unit{fb}), depending on the mass of \PX and \PY. In the \XYggHtt search, the observed (expected) upper limits on the product of the production cross section and \Ygg branching fraction vary between 0.69--15\unit{fb} (0.73--8.3\unit{fb}) in the low \PY mass search and between 0.64--10\unit{fb} (0.70--7.6\unit{fb}) in the high \PY mass search. In the low \PY mass search, these limits are compared to a set of maximally allowed cross sections in the next-to-minimal supersymmetric SM taken from Ref.~\cite{Ellwanger:2022jtd}. A region in the (\mx, \my) parameter space where the observed limits are lower than the maximum allowed is found, meaning that these results can be used to provide tighter constraints on the next-to-minimal supersymmetric SM.

\begin{acknowledgments}
We congratulate our colleagues in the CERN accelerator departments for the excellent performance of the LHC and thank the technical and administrative staffs at CERN and at other CMS institutes for their contributions to the success of the CMS effort. In addition, we gratefully acknowledge the computing centers and personnel of the Worldwide LHC Computing Grid and other centers for delivering so effectively the computing infrastructure essential to our analyses. Finally, we acknowledge the enduring support for the construction and operation of the LHC, the CMS detector, and the supporting computing infrastructure provided by the following funding agencies: SC (Armenia), BMBWF and FWF (Austria); FNRS and FWO (Belgium); CNPq, CAPES, FAPERJ, FAPERGS, and FAPESP (Brazil); MES and BNSF (Bulgaria); CERN; CAS, MoST, and NSFC (China); MINCIENCIAS (Colombia); MSES and CSF (Croatia); RIF (Cyprus); SENESCYT (Ecuador); ERC PRG, RVTT3 and MoER TK202 (Estonia); Academy of Finland, MEC, and HIP (Finland); CEA and CNRS/IN2P3 (France); SRNSF (Georgia); BMBF, DFG, and HGF (Germany); GSRI (Greece); NKFIH (Hungary); DAE and DST (India); IPM (Iran); SFI (Ireland); INFN (Italy); MSIT and NRF (Republic of Korea); MES (Latvia); LMTLT (Lithuania); MOE and UM (Malaysia); BUAP, CINVESTAV, CONACYT, LNS, SEP, and UASLP-FAI (Mexico); MOS (Montenegro); MBIE (New Zealand); PAEC (Pakistan); MES and NSC (Poland); FCT (Portugal); MESTD (Serbia); MICIU/AEI and PCTI (Spain); MOSTR (Sri Lanka); Swiss Funding Agencies (Switzerland); MST (Taipei); MHESI and NSTDA (Thailand); TUBITAK and TENMAK (T\"{u}rkiye); NASU (Ukraine); STFC (United Kingdom); DOE and NSF (USA).

\hyphenation{Rachada-pisek} Individuals have received support from the Marie-Curie program and the European Research Council and Horizon 2020 Grant, contract Nos.\ 675440, 724704, 752730, 758316, 765710, 824093, 101115353, 101002207, 101001205, and COST Action CA16108 (European Union); the Leventis Foundation; the Alfred P.\ Sloan Foundation; the Alexander von Humboldt Foundation; the Science Committee, project no. 22rl-037 (Armenia); the Fonds pour la Formation \`a la Recherche dans l'Industrie et dans l'Agriculture (FRIA-Belgium); the Beijing Municipal Science \& Technology Commission, No. Z191100007219010, the Fundamental Research Funds for the Central Universities, the Ministry of Science and Technology of China under Grant No. 2023YFA1605804, and the Natural Science Foundation of China under Grant No. 12061141002 (China); the Ministry of Education, Youth and Sports (MEYS) of the Czech Republic; the Shota Rustaveli National Science Foundation, grant FR-22-985 (Georgia); the Deutsche Forschungsgemeinschaft (DFG), among others, under Germany's Excellence Strategy -- EXC 2121 ``Quantum Universe" -- 390833306, and under project number 400140256 - GRK2497; the Hellenic Foundation for Research and Innovation (HFRI), Project Number 2288 (Greece); the Hungarian Academy of Sciences, the New National Excellence Program - \'UNKP, the NKFIH research grants K 131991, K 133046, K 138136, K 143460, K 143477, K 146913, K 146914, K 147048, 2020-2.2.1-ED-2021-00181, TKP2021-NKTA-64, and 2021-4.1.2-NEMZ\_KI-2024-00036 (Hungary); the Council of Science and Industrial Research, India; ICSC -- National Research Center for High Performance Computing, Big Data and Quantum Computing, FAIR -- Future Artificial Intelligence Research, and CUP I53D23001070006 (Mission 4 Component 1), funded by the NextGenerationEU program (Italy); the Latvian Council of Science; the Ministry of Education and Science, project no. 2022/WK/14, and the National Science Center, contracts Opus 2021/41/B/ST2/01369 and 2021/43/B/ST2/01552 (Poland); the Funda\c{c}\~ao para a Ci\^encia e a Tecnologia, grant CEECIND/01334/2018 (Portugal); the National Priorities Research Program by Qatar National Research Fund; MICIU/AEI/10.13039/501100011033, ERDF/EU, "European Union NextGenerationEU/PRTR", and Programa Severo Ochoa del Principado de Asturias (Spain); the Chulalongkorn Academic into Its 2nd Century Project Advancement Project, and the National Science, Research and Innovation Fund via the Program Management Unit for Human Resources \& Institutional Development, Research and Innovation, grant B39G680009 (Thailand); the Kavli Foundation; the Nvidia Corporation; the SuperMicro Corporation; the Welch Foundation, contract C-1845; and the Weston Havens Foundation (USA).
\end{acknowledgments}

\bibliography{auto_generated}

\cleardoublepage \appendix\section{The CMS Collaboration \label{app:collab}}\begin{sloppypar}\hyphenpenalty=5000\widowpenalty=500\clubpenalty=5000
\cmsinstitute{Yerevan Physics Institute, Yerevan, Armenia}
{\tolerance=6000
A.~Hayrapetyan, A.~Tumasyan\cmsAuthorMark{1}\cmsorcid{0009-0000-0684-6742}
\par}
\cmsinstitute{Institut f\"{u}r Hochenergiephysik, Vienna, Austria}
{\tolerance=6000
W.~Adam\cmsorcid{0000-0001-9099-4341}, J.W.~Andrejkovic, L.~Benato\cmsorcid{0000-0001-5135-7489}, T.~Bergauer\cmsorcid{0000-0002-5786-0293}, S.~Chatterjee\cmsorcid{0000-0003-2660-0349}, K.~Damanakis\cmsorcid{0000-0001-5389-2872}, M.~Dragicevic\cmsorcid{0000-0003-1967-6783}, P.S.~Hussain\cmsorcid{0000-0002-4825-5278}, M.~Jeitler\cmsAuthorMark{2}\cmsorcid{0000-0002-5141-9560}, N.~Krammer\cmsorcid{0000-0002-0548-0985}, A.~Li\cmsorcid{0000-0002-4547-116X}, D.~Liko\cmsorcid{0000-0002-3380-473X}, I.~Mikulec\cmsorcid{0000-0003-0385-2746}, J.~Schieck\cmsAuthorMark{2}\cmsorcid{0000-0002-1058-8093}, R.~Sch\"{o}fbeck\cmsAuthorMark{2}\cmsorcid{0000-0002-2332-8784}, D.~Schwarz\cmsorcid{0000-0002-3821-7331}, M.~Sonawane\cmsorcid{0000-0003-0510-7010}, W.~Waltenberger\cmsorcid{0000-0002-6215-7228}, C.-E.~Wulz\cmsAuthorMark{2}\cmsorcid{0000-0001-9226-5812}
% ...
\par}
\vskip\cmsinstskip
\dag:~Deceased\\
$^{1}$Also at Yerevan State University, Yerevan, Armenia\\
$^{2}$Also at TU Wien, Vienna, Austria\\
% ...
\end{sloppypar}
\end{document}